\documentclass[twocolumn]{aastex63}
\usepackage{natbib}
\newcommand{\squiggle}{SQuIGG$\vec{L}$E \,}
\newcommand{\squigglecomma}{SQuIGG$\vec{L}$E}

\newcommand{\sersic}{S\'{e}rsic \,}

\usepackage{amsmath} % or simply amstext

\usepackage[encapsulated]{CJK}

\usepackage{rotating}

\usepackage{float}
\usepackage{textcomp}
\usepackage{makecell}

\begin{document}

\date{\today}

% \submitjournal{ApJ Letters}

\shorttitle{Young Post-Starburst Galaxies are CO-Luminous}

\shortauthors{Setton et al.}

\title{SQuIGG$\vec{L}$E: Buried star formation cannot explain the rapidly fading CO(2-1) luminosity in massive, $z\sim0.7$ post-starburst galaxies}

%core authors
\author[0000-0003-4075-7393]{David J. Setton}\thanks{Email: davidsetton@princeton.edu}\thanks{Brinson Prize Fellow}
\affiliation{Department of Astrophysical Sciences, Princeton University, 4 Ivy Lane, Princeton, NJ 08544, USA}

\author[0000-0003-3256-5615]{Justin~S.~Spilker}
\affiliation{Department of Physics and Astronomy and George P. and Cynthia Woods Mitchell Institute for Fundamental Physics and Astronomy, Texas A\&M University, 4242 TAMU, College Station, TX 77843-4242, US}

\author[0000-0001-5063-8254]{Rachel Bezanson}
\affiliation{Department of Physics and Astronomy and PITT PACC, University of Pittsburgh, Pittsburgh, PA 15260, USA}

\author[0000-0002-1714-1905]{Katherine A. Suess}
\affiliation{Department for Astrophysical \& Planetary Science, University of Colorado, Boulder, CO 80309, USA}

\author[0000-0002-5612-3427]{Jenny E. Greene}
\affiliation{Department of Astrophysical Sciences, Princeton University, 4 Ivy Lane, Princeton, NJ 08544, USA}

\author[0000-0003-4700-663X]{Andy~D.~Goulding}
\affiliation{Department of Astrophysical Sciences, Princeton University, 4 Ivy Lane, Princeton, NJ 08544, USA}

\author[0000-0002-0766-1704]{Elia Cenci}
\affiliation{Department of Astronomy, University of Geneva, Chemin Pegasi 51, Versoix CH-1290, Switzerland}
\affiliation{Department of Astrophysics, Universit\"at Z\"urich, Winterthurerstrasse 190, Zurich CH-8057, Switzerland}

\author[0000-0002-1759-6205]{Vincenzo~R.~D'Onofrio}
\affiliation{Department of Physics and Astronomy and George P. and Cynthia Woods Mitchell Institute for Fundamental Physics and Astronomy, Texas A\&M University, 4242 TAMU, College Station, TX 77843-4242, US}

\author[0000-0002-1109-1919]{Robert~Feldmann}
\affiliation{Department of Astrophysics, Universit\"at Z\"urich, Winterthurerstrasse 190, Zurich CH-8057, Switzerland}

\author[0000-0002-7613-9872]{Mariska~Kriek}
\affiliation{Leiden Observatory, Leiden University, P.O. Box 9513, 2300 RA Leiden, The Netherlands}

\author[0009-0005-4226-0964]{Anika Kumar} 
\affiliation{Laboratory for Multiwavelength Astrophysics, School of Physics and Astronomy, Rochester Institute of Technology, 84 Lomb Memorial Drive, Rochester, NY 14623, USA} 
\affiliation{Department of Physics and Astronomy and PITT PACC, University of Pittsburgh, Pittsburgh, PA 15260, USA} 

\author[0000-0002-0696-6952]{Yuanze~Luo}
\affiliation{Department of Physics and Astronomy and George P. and Cynthia Woods Mitchell Institute for Fundamental Physics and Astronomy, Texas A\&M University, 4242 TAMU, College Station, TX 77843-4242, US}

\author[0000-0002-7064-4309]{Desika~Narayanan}
\affiliation{Department of Astronomy, University of Florida, 211 Bryant Space Science Center, Gainesville, FL 32611, USA}

\author[0000-0003-1535-4277]{Margaret E. Verrico}
\affiliation{University of Illinois Urbana-Champaign Department of Astronomy, University of Illinois, 1002 W. Green St., Urbana, IL 61801, USA}
\affiliation{Center for AstroPhysical Surveys, National Center for Supercomputing Applications, 1205 West Clark Street, Urbana, IL 61801, USA}

\author[0000-0002-6768-8335]{Pengpei~Zhu}
\affiliation{Cosmic Dawn Center (DAWN), Denmark}
\affiliation{DTU Space, Technical University of Denmark, Elektrovej 327, 2800 Kgs. Lyngby, Denmark}

\submitjournal{ApJ}

\begin{abstract}

Observational and theoretical studies have long held that rapid gas consumption in starbursts is responsible for the formation of quiescent galaxies. However, studies of recently quenched ``post-starburst" galaxies have discovered that a number of them are surprisingly luminous in CO, challenging this assumption. We present deep ALMA CO(2-1) observations of 50 massive ($\log(M_\star/M_\odot)\sim11.2$) post-starburst galaxies from the \squiggle sample at $z\sim0.7$. We detect a large fraction (27/50) of the galaxies in CO(2-1). Furthermore, we find that the CO luminosity correlates with the age of the recent starburst, suggesting a gas-removal timescale of $\lesssim140$ Myr, an order of magnitude shorter than is implied by their rest optical star formation rates. We perform new spectral energy distribution fits incorporating mid- and far-IR photometry to test whether dust-obscured star formation can explain this trend. We find that while allowing for buried star formation can raise star formation rates by $\sim0.5$ dex, for almost all galaxies it is neither required to fit the observed IR SED, nor is it sufficient to explain the observed depletion trend. Even the combination of significant buried star formation \textit{and} ULIRG-like $\alpha_{CO}$ is not enough to explain this decay in CO luminosity. Furthermore, there is no strong evidence to support either of those modifications to the depletion time. Therefore, it remains a distinct possibility that the age-CO luminosity trend should not be interpreted as an evolutionary sequence, and that gas-rich \squiggle galaxies will soon rejuvenate. 
\end{abstract}

% \keywords{High-redshift galaxies (734); Galaxy quenching (2040); Galaxy evolution (594); Quenched galaxies (2016); Post-starburst galaxies (2176)}

%%%%%%%%%%%%%%%%%%%%%%%%%%%%%%%%%%%%%%%%%%%%%%%%%%%%%%%%%%%%%%%%%%%%%%%%%%%%%%%%%%%%%%%%%%%%%%%%%%%%%%%%%%%%%%%%%%%%%%%%%%%%%%%%%%%%%%%%%%%%%%%%%%%%%%%%%%%%%%%%%%%%%%%%%%%%%%%%%%%%%%%%%%%%%%%%%%%%%%%%%%%%%%%%%%%%%%%%%%%%%%%%%%%%%%%%%%%%%%%%%%%%%%%%%%

\section{Introduction} \label{sec:intro}

One of the largest unsolved questions in galaxy evolution is the physical cause of the cessation of star formation--or ``quenching"--in the most massive galaxies. The observed bi-modality in galaxy color \citep[e.g,.][]{Shen2003} and morphology \citep[e.g.,][]{VanDerWel2014} suggests that blue, star-forming disks and red, quiescent elliptical galaxies constitute different populations. This implies that the transformation of morphology must have accompanied the cessation of star formation. Additionally, this implies that once galaxies are ``red and dead," they must continue to suppress star formation to remain so, likely due to the influence of active galactic nucleus (AGN) feedback \citep{Hopkins2006, Croton2006}. 

A clear feature that has emerged in the study of large galaxy populations is the presence of a so-called ``green valley," a gap between star-forming galaxies and the quiescent populations that emerges in color-magnitude diagrams and the star forming main sequence (SFMS). The sparse population of this region, especially at high masses, indicates that galaxies spend very little time at intermediate age and star formation rates, and instead quench rapidly \citep[e.g.,][]{Schawinski2014}. This has led to paradigm of two quenching pathways, with a slow mode dominating at low-redshift and a rapid mode dominating in the early universe \citep[e.g.,][]{Belli2019}. Far-IR evidence of a hidden population of dust-obscured star-forming galaxies in the valley can allow for smoother evolution that diminishes the importance of the rapid quenching track \citep[e.g.,][]{Eales2018}. However, the implied short formation timescale of the most massive quiescent systems from abundance and star formation history measurements, both in the oldest galaxies in the local universe \citep[e.g.,][]{Thomas2005, Greene2013, McDermid2015, FerreMateu2017} and in evolved systems at and above cosmic noon \citep[e.g.,][]{Kriek2016,Glazebrook2017,Carnall2019,Man2021,Antwi-Danso2025,Carnall2023a, Glazebrook2024,Beverage2024,deGraaff2025_QG, Carnall2024,Beverage2025, Zhang2025_QG_numberdensity,McConachie2025}, does suggest that a rapid quenching pathway is important in the formation of massive galaxies at high-redshift.

Observationally, one can study the rapid quenching process via its immediate descendants, post-starburst galaxies \citep{Dressler1983, Zabludoff1996}. These galaxies exhibit deep Balmer absorption features, indicating that that their stellar populations are dominated by $\sim$a few hundred Myr old stellar populations as the result of a rapid decline in their star formation rate. Post-starburst galaxies are exceedingly rare in the local universe \citep[e.g.,][]{Wild2009, Pattarakijwanich2016}. However, their rising number density with redshift in conjunction with the falling number density of older quiescent galaxies \citep[e.g.,][]{Whitaker2012a,Wild2016, Rowlands2018a, Belli2019, Park2022,Setton2023, Park2024, Adscheid2025} indeed suggests that they represent the evolutionary link between rapid quenching and massive elliptical galaxies. As the direct products of rapid quenching, studying post-starburst galaxies can illuminate the physical causes of the rapid quenching process.

Perhaps the most important physical pieces of the puzzle of rapid quenching is the state of the molecular gas that fuels star formation. Massive, evolved quiescent galaxies are predominantly gas poor, and have very low molecular gas fractions ($M_\mathrm{gas}/M_\star$) relative to co-eval star forming systems, whether in CO and dust continuum (the best extragalactic tracers of the molecular gas mass) surveys and stacking experiments in the local volume \citep[e.g.,][]{Saintonge2011,Young2011, Davis2016, Michalowski2019, Michalowski2024} or at high-z \citep{Sargent2015,Spilker2018, Williams2021, Whitaker2021a, Adscheid2025}. While there is indirect evidence from stacking analysis of dust continuum and $A_V$ measurements that typical quiescent systems at high-z may be more ISM-rich than their low-z counterparts \citep[e.g.,][]{Gobat2018, Martis2019, Akhshik2023, Setton2024_uncover, Lee2024, Siegel2025}, the bimodality in gas content across cosmic time implies that depletion of molecular gas reservoirs is a prerequisite to quenching. As such, theoretical models have largely sought to explain the mechanisms by which galaxies halt additional accretion from the cosmic web \citep[e.g.,][]{Keres2005,Dekel2006,Feldmann2015}. In these models, the majority of the molecular gas that fueled the burst of star formation prior to quenching would be consumed in the burst and driven out by the combination of quasar- and star-formation-driven outflows, and by the time galaxies are observed as post-starburst, they should already be gas poor \citep[e.g.,][]{Hopkins2008a}.

It has therefore been surprising that follow-up observations of young quiescent galaxies, in contrast with more evolved quiescent systems, have found that most are CO luminous, implying molecular gas fractions as high as 20\% in systems that appear to lie below the star forming main sequence \citep{French2015, Rowlands2015, Alatalo2016b, Suess2017, Yesuf2017, Smercina2018, French2018a, Belli2021, Bezanson2022a, Woodrum2022,Otter2022,Wu2023,Baron2023,Michalowski2024,Umehata2025,Suess2025}. Perhaps even more surprising, the molecular gas fraction is strongly correlated with the post-burst age, such that the youngest post-starburst galaxies are the most CO-luminous. This implies that these gas reservoirs are being removed on short ($\sim$hundreds of Myr) timescales in a way that is incommensurate with the star formation rate, at least as measured by rest-optical tracers \citep{French2015, Bezanson2022a}.

There have been numerous proposed explanations for the state of this molecular gas, and for the ultimate evolutionary pathway, of these post-starburst systems. Ionized \citep{Tremonti2007,Sell2014,Baron2018,Baron2020,Diamond-Stanic2021, Fodor2025}, neutral \citep{Alatalo2016a,Baron2020,Baron2022a,Luo2022,Belli2023, Park2024, Sun2025_PSBoutflow, Wu2025, Valentino2025}, and molecular \citep{Geach2014, Geach2018, Spilker2020a, Spilker2020b} outflows have been identified in recent starburst and post-starburst systems, but it is unclear whether these starburst or AGN driven winds can remove the bulk of the molecular gas seen in these systems. The CO reservoirs of post-starburst galaxies have been found to be morphologically disturbed \citep{Smercina2022, Sun2022}, and in some systems as much at 50\% of the CO luminosity has been found on extended scales of tens of kiloparsecs along tidal features \citep{Spilker2022, Donofrio2025}, suggesting that merger driven disruption may play a role in removing gas and suppressing star formation. Together, the injection of energy by the supernovae, AGN, and mergers that drive this removal and the dynamical effects of the compact post-starburst cores \citep{Almaini2017,Maltby2018,Wu2020,Diamond-Stanic2021,Setton2022,Zhang2024} may play a role in morphologically quenching the remaining gas and stopping it from forming new stars \citep[e.g.,][]{Martig2009}.

However, the finding that many gas-rich post-starburst galaxies are mid-IR luminous \citep[e.g.,][]{Alatalo2017, Yesuf2017, Smercina2018,Baron2022a, Suess2022a} has also promoted the potential interpretation that these CO-luminous galaxies that present in the rest-optical as post-starburst are actually still in the midst of their bursts, hiding deeply obscured starbursts behind optically thick dust \citep[e.g.,][]{Smail1999,Poggianti2000,Baron2023}. In this physical picture, the CO luminosity can be explained as the fuel for these active starbursts, and the depletion time of the molecular gas can be commensurate with the observed fading reservoirs on timescales of a few hundred Myr \citep{French2015, Bezanson2022a}.

In this work, we set out to test the aforementioned claims, using the \squiggle \citep[Studying Quenching in Intermediate-z Galaxies: Gas, angu$\vec{L}$ar momentum, and Evolution, see][]{Suess2022a} sample, the largest spectroscopic Sample of massive post-starburst galaxies at $z=0.5-0.9$. By combining new and ancillary CO(2-1) and dust continuum observations, quadrupling the total sample size, with rest-optical spectroscopy and rest-optical-to-far-infrared photometry from the Sloan Digital Sky Survey DR14 \citep[SDSS,][]{Abolfathi2018}, the Wide Infrared Survey Explorer \citep[WISE, ][]{Wright2010}, and the Herschel Space Observatory \citep{Pilbratt2010, Poglitsch2010, Griffin2010}, we perform new spectrophotometric fitting to quantify the level of allowable buried star formation in these systems and to test whether it can explain their over-luminous CO reservoirs. This work is laid out as follows. In Section \ref{sec:data}, we present the \squiggle sample and describe our new Atacama Large Millimeter Array (ALMA) CO(2-1) and dust continuum observations to assemble the largest sample of molecular gas measurements of post-starburst galaxies to date. In Section \ref{sec:analysis}, we contextualize our CO(2-1) measurements and present new spectrophotometric fits, building on the detailed \texttt{Prospector} \citep{Johnson2017,Leja2017, Johnson2021} modeling presented \cite{Suess2022a} via the inclusion of mid- and far-IR constraints. In Section \ref{sec:results}, we analyze how these new star formation histories impact our conclusions about the star formation rate, the IR luminosity, and the depletion time of these systems. In Section \ref{sec:tdep}, we discuss the implications of these star formation histories on the depletion time in detail. Finally, in Section \ref{sec:conclusions}, we summarize our results. In an accompanying research note (Kumar et al. submitted), we also characterize the serendipitously CO-detected ``buddy" galaxies that are likely satellites of the \squiggle host galaxies to constrain the environments of these post-starburst galaxies. 

Throughout this work, we adopt the best-fit cosmological parameters from the WMAP 9 year results \citep{Hinshaw2013}: $H_0 = 69.32 \ \mathrm{km \ s^{-1} \ Mpc^{-1}}$, $\Omega_m = 0.2865$, and $\Omega_\Lambda = 0.7135$, utilize a Chabrier initial mass function \citep{Chabrier2003}, and quote AB magnitudes. For convenience, throughout the work we quote $\alpha_{CO}$ without its units of $M_\odot \ \mathrm{(K \ km \ s^{-1} \ pc^{-2})^{-1}}$.

\section{Data} \label{sec:data}

\subsection{The \squiggle Sample}

The \squiggle Sample was selected from the Sloan Digital Sky Survey DR14 \citep[SDSS,][]{Abolfathi2018} to constitute a pure sample of post-starburst galaxies at $z=0.5-0.9$ based on their spectroscopic shapes. In particular, they were selected using $U_m$, $B_m$, and $V_m$ rest-frame colors defined in \cite{Kriek2010} to have strong Balmer breaks, but also to have blue slopes redward of the break to select against old or dusty populations. A full description of the sample and detailed rest-optical spectrophotometric fitting can be found in \cite{Suess2022a}. In total, the sample consists of 1318 galaxies, with a median redshift of 0.7 and SDSS $i_{AB}=19.5$, corresponding to $\log(M_\star/M_\odot)\sim11.2$.

Initially, twelve of the most luminous and massive \squiggle galaxies (see Figure \ref{fig:sample}) were targeted for follow-up. Spatially resolved rest-optical spectroscopy of these systems with Gemini/GMOS revealed a range of ordered motion and flat age gradients as traced by $H\delta_A$ equivalent width \citep{Hunt2018, Setton2020}. Additionally, deep ($\sim2$ hour/galaxy) ALMA Band 4 observations of these systems revealed that many of them are surprisingly luminous in CO(2-1), suggesting significant residual molecular gas despite their apparent quiescence as determined by rest-optical spectrophotometric fitting \citep{Suess2017, Bezanson2022a}. While much of this CO(2-1) appears bound to the galaxies, a significant amount (as high as $\sim50\%$) of the luminosity is found to emit at distances of 10s of kiloparsecs along tidal tails that are revealed in HST imaging and high resolution ALMA observations \citep{Spilker2022, Donofrio2025}. All 12 of the CO(2-1)-observed galaxies were targeted for rest-optical spectroscopy of $H\alpha$, with weak emission confirming their low star formation rates, albeit with systematic uncertainty in the dust correction due to the inability to correct for Balmer decrement \citep{Zhu2025}.

There is an additional subset of \squiggle that is distinguished from the rest of the sample by its overlap with the Hyper-Suprime Cam imaging survey \citep[HSC][]{Aihara2018}, enabling the study of their structures. \cite{Setton2022} performed \sersic fitting of these 145 galaxies and found that they lie systematically below the mass-size relation, even relative to other quiescent galaxies, suggesting that their proto-elliptical structure is in place at the time of quenching, and \cite{Verrico2023} showed that they host a very high incidence of tidal features, especially in the youngest systems, suggesting a link between their burst and quenching and major mergers.

\subsection{ALMA observations}

\begin{figure}
    \centering
    \includegraphics[width=0.49\textwidth]{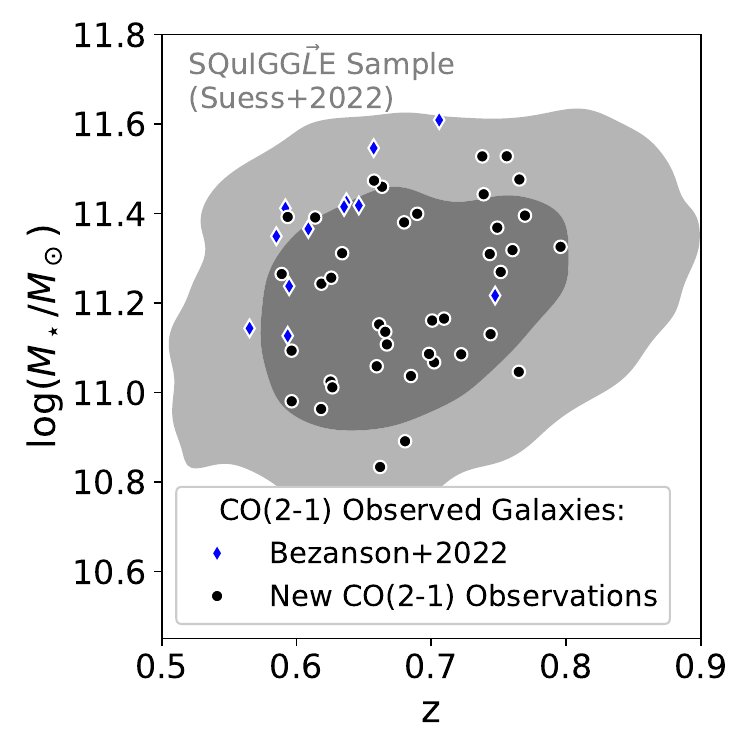}
    \caption{\squiggle is an SDSS-selected sample of 1318 massive ($\log(M_\star/M_\odot) \sim 11.2$) post-starburst galaxies at $z\sim0.7$ \citep[grey contours,][]{Suess2022a}. Previous observational work observing CO(2-1) in these galaxies preferentially targeted the brightest (highest mass/lowest-z) galaxies \citep[blue points,][]{Suess2017, Bezanson2022a}. In this work, we quadruple our sample size and target a much more representative sample (black points).}
    \label{fig:sample}
\end{figure}

\begin{figure*}
    \centering
    \includegraphics[width=\textwidth]{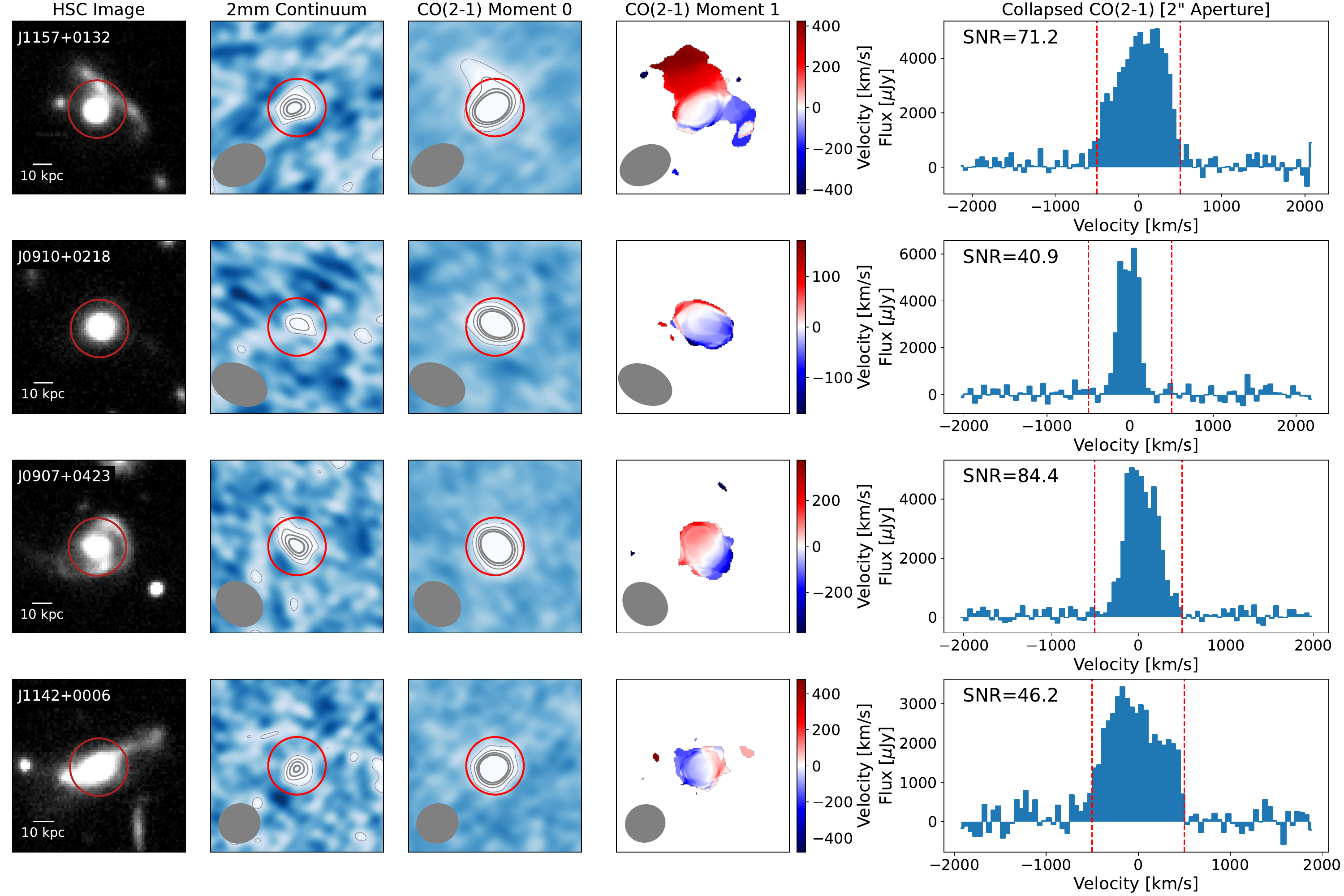}
    \caption{$12''\times12''$ cutouts of the four most CO-luminous \squiggle galaxies in the new sample. For each galaxy, we show the HSC-i band image, the 2mm continuum image (collapsing the three ALMA spectral windows that were not centered on CO(2-1), the CO(2-1) Moment 0 image, the CO(2-1) Moment 1 image, and the collapsed CO(2-1) spectrum within a 2" aperture (also indicated as a red circle). In all ALMA images, the synthesized beam is shown as a grey circle, and contours of 2, 4, 6, and 8$\sigma$ are shown as increasingly thick lines. The velocity region integrated to measure the CO flux is shown with dashed lines.}
    \label{fig:most_gas_rich}
\end{figure*}

\begin{figure*}
    \centering
    \includegraphics[width=\textwidth]{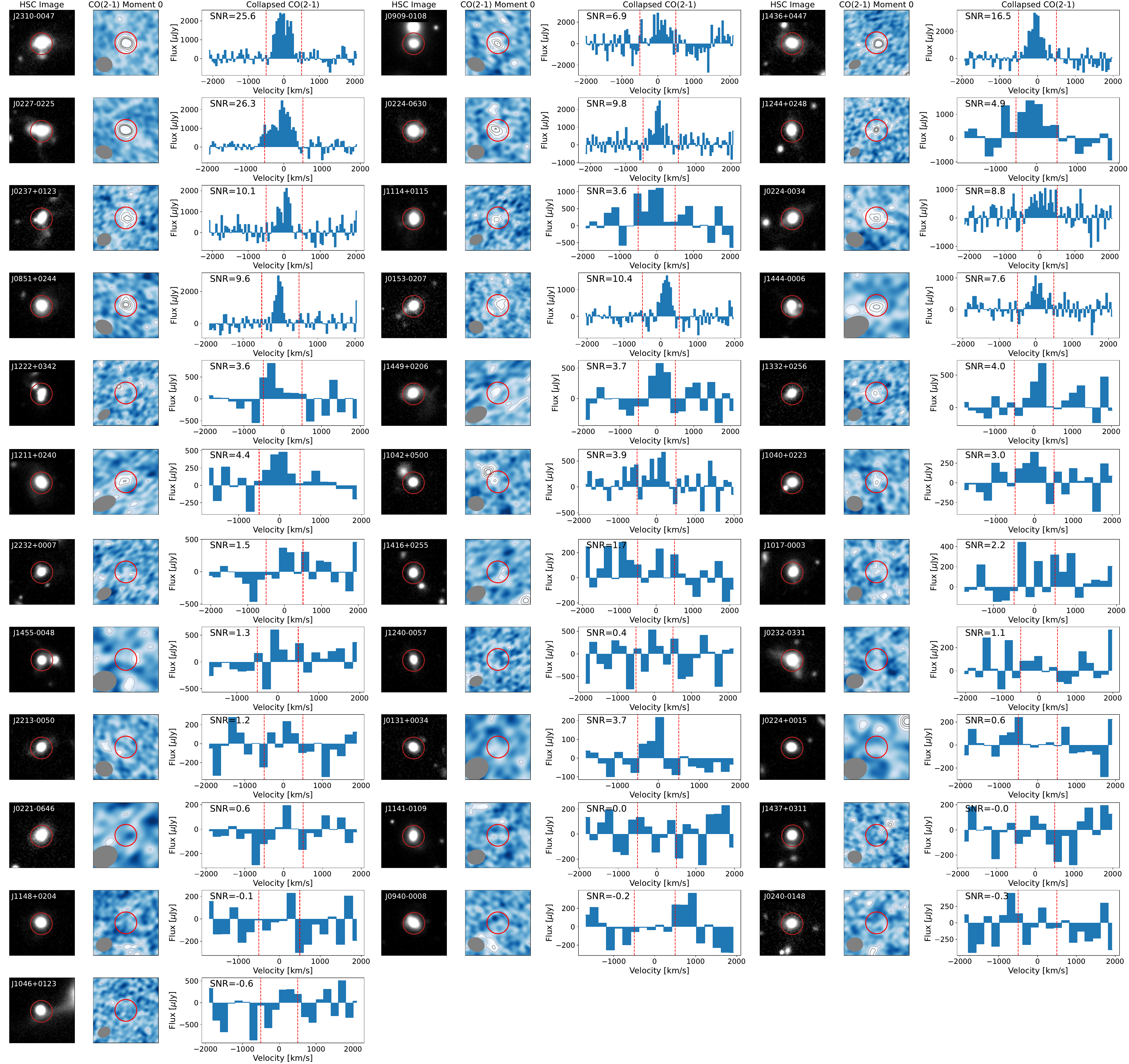}
    \caption{As in Figure \ref{fig:most_gas_rich}, but showing only the HSC image, the CO(2-1) Moment 0 map, and the collapsed CO(2-1) spectrum for the entire rest of the new galaxies (sorted by CO(2-1) luminosity) presented in this work.}
    \label{fig:rest_of_sample}
\end{figure*}

In this work, we compile a range of Band 4 observations of \squiggle galaxies to expand on previous analysis of CO(2-1) and dust continuum measurements, quadrupling our sample relative to \cite{Bezanson2022a}. Furthermore, this new is far more representative sample of \squiggle, as galaxies were selected to span a range of key parameters \citep[provided they overlapped with HST imaging, see][]{Setton2022}, in contrast with the \cite{Bezanson2022a} sample that drew from the brightest sources. Here, we describe those programs, all of which were observed in compact configurations as detection experiments with beam sizes ranging between 0.7-2.2$''$.

The majority of the new data presented in this work comes from Program 2021.1.01535.S and its re-submission, 2022.1.00604.S (PI: D. Setton). These observations, which were taken at various times throughout 2022 and 2023, were part of a survey program that targeted a sample of 31 galaxies spanning a range in stellar mass, star formation rate, and age \citep[as determined by the fits in][]{Suess2022a}, in addition to redshift and rest-optical size \citep[as measured in][]{Setton2022, Verrico2023}. Observations in this program were targeted for 1 hour, although a small number of targets that were observed between the Cycle 9 submission deadlines and the end of Cycle 8 were targeted twice, resulting in a total integration time closer to two hours.

An additional six galaxies were targeted as a part of Cycle 8 Program 2021.1.00988.S (PI: D. Setton). These galaxies were specifically targeted as young mergers \citep[via their tidal features, see][]{Verrico2023}, similar to those presented in \citep{Spilker2022, Donofrio2025}, with the goal of identifying tidally stripped gas. As such, galaxies in this program were observed for $\sim2$ hours/galaxy to mimic the depths of the galaxies presented in \cite{Suess2017} and \cite{Bezanson2022a} where the spatially-extended CO was first identified. Because these galaxies were specifically targeted to select for CO luminous sources, they were detected at a high rate (5/6) and are among the most CO luminous sources in our sample. Finally, one galaxy was observed as a part of the partially completed Cycle 8 Program 2021.1.00761.S (PI: K. Suess), which targeted AGN candidates identified in \cite{Greene2020} that overlapped with the HSC footprint. Details of the targets and observing strategy are presented in Table \ref{tbl:alma_properties}.

In Figure \ref{fig:sample}, we show the distribution of these new observations in $\log(M_\star/M_\odot)$ versus redshift as black points, as compared to the full \squiggle sample (grey contours) and the \citep{Bezanson2022a} sample (blue diamonds). In contrast with the previous observations, which targeted a subset that was heavily biased toward lower-redshift massive sources, this new sample does a much better job at spanning the full range of \squiggle galaxies. While we cannot purport that this sample is ``complete" (as the \squiggle sample itself is a heavily biased subset of SDSS galaxies, largely drawn from the BOSS CMASS sample, but also spanning a wide range of other selection criteria), it will at least span a wide range in the properties of galaxies that meet the rest-frame selection criteria that \cite{Suess2022a} demonstrated result in a pure sample of galaxies that just underwent a rapid starburst, followed by a precipitous drop in star formation rate. In the next two sections, we describe our reduction of the ALMA data, the synthesis of CO(2-1) and continuum cubes, and our extraction of line and continuum fluxes. All imaging was performed with CASA version 6.6.3-22 \citep{CASA2022}.

\subsubsection{CO maps and flux measurements}

We image all galaxies in the channel centered on CO(2-1) with 0.08$''$ cells, using natural weighting and combining all available UV data. We utilize the \texttt{tclean} algorithm with the \texttt{auto-multithresh} algorithm, setting \texttt{pblimit=-0.1} and \texttt{nsigma=3}. For each galaxy, we generate maps at 50, 100, and 200 km/s, centered at the systemic velocity of our galaxy. We then extract 1D spectra using a 2$''$ aperture at the location of our galaxy. We measure the total CO flux by integrating at $\pm$500 km/s, estimating uncertainties using the rms of the spectrum outside this region. All measurements are first made on the 50 km/s resolution cube; if the galaxy is not detected at the 3$\sigma$ level, we then iteratively use the 100 km/s and 200 km/s cubes, marking a galaxy as a non-detection if it is not significantly detected in any of the 3 cubes. The measured CO flux densities and luminosities are presented in Table \ref{tbl:alma_measurements}. In Figure \ref{fig:most_gas_rich} and \ref{fig:rest_of_sample}, we show the Moment 0 maps in addition to the collapsed CO(2-1) spectra, with the velocity channel used for integration indicated on the spectra.

For illustrative purposes, we also generate Moment 1 maps for our four most CO-luminous sources, which we obtain by performing a velocity-weighted average over the CO luminosity within $\pm500$ km/s. All four of these galaxies show some degree of velocity gradients, with disrupted kinematics that often mirror the tidal features seen in the rest-optical HSC imaging. However, as our resolution is low and the sources are only marginally resolved, we defer any detailed kinematic and structural analysis to future work. In Figure \ref{fig:rest_of_sample}, we show the rest of the new galaxies observed in this program, restricting to only the HSC image, CO(2-1) moment 0 map, and extracted CO spectrum.

\subsubsection{2 mm continuum measurements} 
\label{subsec:2mm}

We image the continuum with identical \texttt{tclean} parameters to the previous section, using the three spectral windows that do not contain CO(2-1) emission and \texttt{specmode=`mfs'}. We also measure fluxes within a $2''$ aperture, estimating uncertainties with the RMS of the image under the assumption that our sources are unresolved. In Figure \ref{fig:most_gas_rich}, we show the 2mm continuum images for our four most luminous sources from the new sample, all of which are detected. However, only 11/50 galaxies in the full \squiggle galaxies are continuum detected at the 3$\sigma$ level. The measured continuum flux densities are presented in Table \ref{tbl:alma_measurements}.

\subsubsection{Herschel photometry}

In order to further supplement our FIR constraints near the peak of the dust SED, we search for archival Herschel Space Observatory \citep{Pilbratt2010, Poglitsch2010, Griffin2010} photometry of the \squiggle targets in our survey. Specifically, we query both the PACS and SPIRE point source catalogs \citep{Marton2017,Schulz2017} at the locations of our sources, searching all bands for matches where the centroid in the point source catalog is within 3$''$ of our source. There are no PACS 70$\mu$m, 100$\mu$m, or 160$\mu$m matches for any of our sources that do not also appear in the Rejected Source List. Similarly, no sources appear in the SPIRE 350 $\mu$m or 500 $\mu$m catalogs. However, we find three sources (J1157+0132, J0910+0218, and J1142+0006) with matches in the 250 $
\mu$m catalog, all only barely detected at 3-4$\sigma$ significance. We include this photometry in our analysis.

\begin{figure*}
    \centering
    \includegraphics[width=\textwidth]{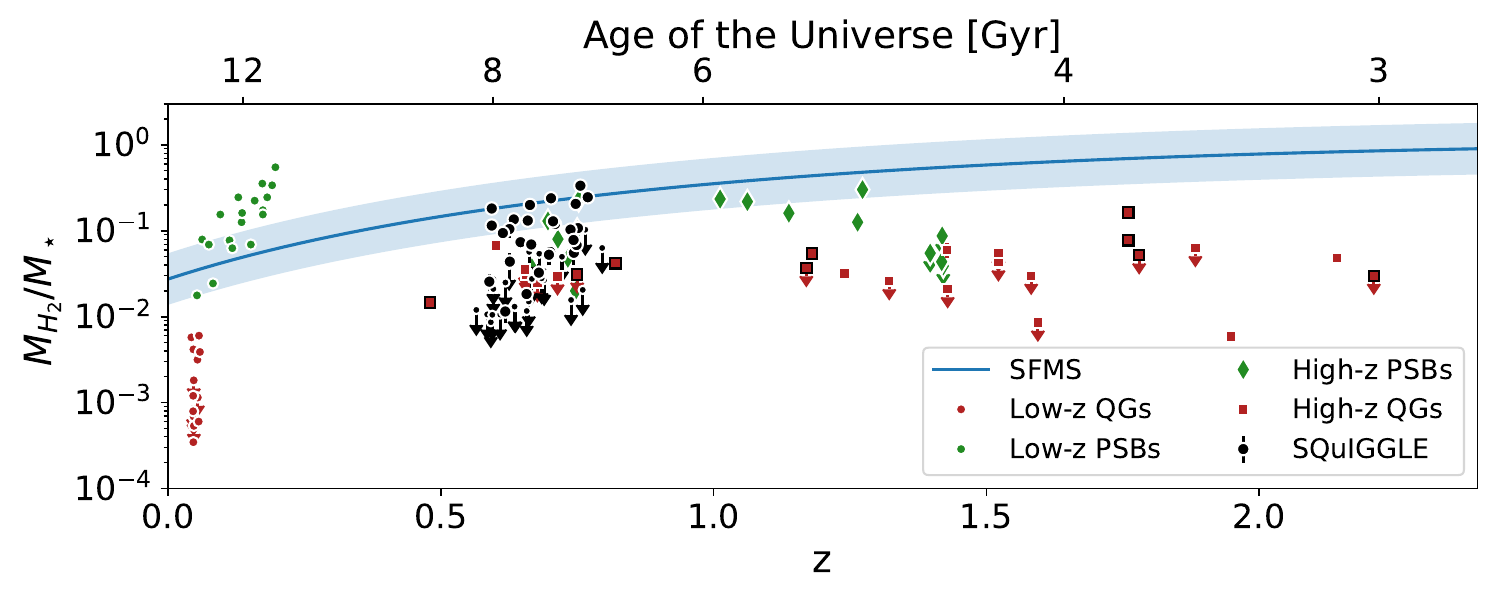}
    \caption{Redshift versus the molecular gas fraction for passive galaxies with $\log(M_\odot/M_\star)>10.8$. We show the scaling relation for star forming galaxies with $\log(M_\star/M_\odot)=11$ from \cite{Tacconi2018} in blue with the shaded region denoting 0.3 dex scatter. Low- \citep{Davis2016} and high-z \citep{Sargent2015,Spilker2018,Gobat2018,Magdis2021,Caliendo2021,Whitaker2021a, Williams2021, Adscheid2025} quiescent galaxies are shown as red circles and squares respectively, with white outlines indicating measurements from individual sources and black outlines indicating measurements from stacking. Similarly, low- and high-z post-starburst galaxies from the literature are shown as green circles \citep{French2015, Alatalo2016b,Baron2023} and diamonds \citep{Spilker2018, Zanella2023,Wu2023, Suess2025}. When literature measurements are of dust continuum, $\delta_{GDR}=100$ is adopted. Finally, as black circles, we show the \squiggle sample. In contrast with older quiescent galaxies, which tend to have very low molecular gas fraction across cosmic time, post-starburst galaxies exhibit a wide spread in molecular gas fraction. This is exemplified by \squigglecomma, where some galaxies have gas fractions high enough to place them on the main sequence and some galaxies have fractions that are constrained to lie more than an order of magnitude below.}
    \label{fig:gas_frac_z}
\end{figure*}

While we do not include non-detections in our analysis (as non-detections at the typical depth of Herschel are not particularly constraining for sources in our redshift range), we do note that while we only found that 3 galaxies were securely detected, 22/50 sources are either detected or have an entry in the point source catalog within 60$''$ (suggesting that they are in the archival imaging footprint of Herschel). As such, we conclude that while a few exceptional sources were detected, the lack of matches is not primarily due to the lack of coverage, and instead it seems that the typical \squiggle galaxy is not IR-luminous enough to be detected in Herschel imaging. 

\section{Analysis} \label{sec:analysis}

In this section, we seek to contextualize the observed CO properties of the \squiggle post-starburst sample within the evolutionary trajectory that can transform star forming galaxies into red-and-dead quiescent galaxies. Doing so relies heavily on our understanding of the stellar populations of these galaxies. \cite{Suess2022a} presented detailed spectrophotometric fits to the rest-optical SED of the full \squiggle sample, finding that these sources are characteristically massive ($\log(M_\star/M_\odot)\sim11.2$) and quiescent (log(sSFR $[\mathrm{yr^{-1}}]) \lesssim -11$). In the first sections, we utilize these fits, in conjunction with our new and previously published \citep{Suess2017, Bezanson2022a} CO(2-1) measurements, to place these sources into context under the standard set of assumptions in that previous work. Then, we shift our focus to the stellar populations of \squiggle themselves, addressing how changing those assumptions and incorporating data from the far-IR SED might change the inferences about the quiescence of these systems. 

\subsection{The molecular content of post-starburst galaxies}

First, we evaluate how our the molecular gas mass fraction of the \squiggle sample compares to the expectation for a population that has finished its primary epoch of star formation and is entering quiescence. To do so, in Figure \ref{fig:gas_frac_z} we compare our sample to scaling relations for the time evolution of the molecular gas fraction for star forming massive ($\log(M_\star/M_\odot)=11$) galaxies from \cite{Tacconi2018}. The molecular gas fraction evolves with cosmic time, as the entire population of galaxies has become less gas-rich and less star forming as the universe has evolved in the wake of cosmic noon. However, across cosmic time, massive quiescent galaxies (shown in red) lie well below this relation, indicating that a large part of why they are no longer star forming is because they no longer host sufficient fuel for substantial new star formation.

Post-starburst galaxies (shown in green), however, which potentially represent the evolutionary link between these populations, show much more mixed properties. While many post-starburst galaxies, especially at high-z, are not detected in CO or dust continuum, finding limits that are consistent with the gas-poor quiescent population, many sources are detected with gas fractions that place them on (or at low-z, above, predominantly in the SPOG sample, see \citealt{Alatalo2016b}) the star forming main sequence at their redshift. Our compiled \squiggle galaxies reflect this trend near cosmic noon. Under the standard assumption of $\alpha_{CO}=4.0$ \citep{Bolatto2013} and $r_{21}=1.0$, we find that our most strongly detected galaxies have molecular gas fractions as high as 30\%, while non-detections are constrained to have fractions as low as $0.3\%$. This finding reflects a broad consensus in the literature that many optically selected post-starburst galaxies at and below cosmic noon can host molecular gas fractions that are typical, if slightly low, relative to co-eval star forming galaxies \citep{Suess2017, Spilker2018, Bezanson2022a,Zanella2023, Wu2023}. However, there also exists a significant population of post-starburst systems, largely probed by \squigglecomma, that fall significantly below the main sequence, with molecular gas fractions that are commensurate with the limits from older high-z quiescent systems \citep[e.g.,][]{Sargent2015, Magdis2021, Whitaker2021a, Williams2021}.

\begin{figure}
    \centering
    \includegraphics[width=0.5\textwidth]{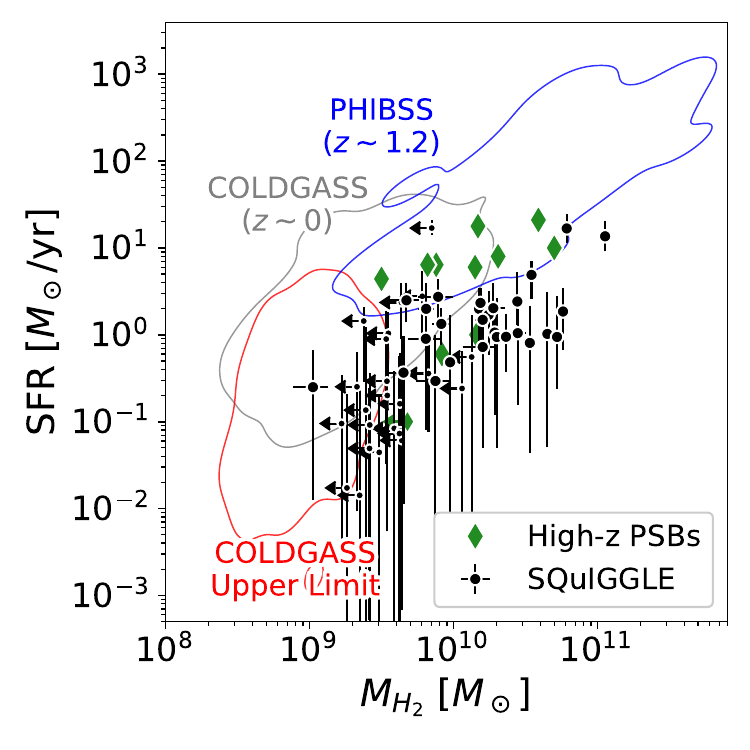}
    \caption{The molecular gas mass versus the star formation rate. As background contours, we show literature star forming samples: COLDGASS detected (grey) undetected \citep[red,][]{Saintonge2011} and  PHIBSS/PHIBSS2 \citep[blue,][]{Tacconi2018,Freundlich2019}. As green diamonds, we show literature post-starburst samples \citep{Belli2019,Woodrum2022, Wu2023}, with star formation rates derived from SED fitting that incorporates mid-IR data. In black, we show \squiggle, with the star formation rates measured in \cite{Suess2022a}. Under the assumptions in that fitting (and with $\alpha_{CO}=4.0$), the entire \squiggle sample lies below the SFR-$M_{H_2}$ relation, suggesting that star formation in \squiggle post-starburst galaxies at fixed $M_{H_2}$ is low relative to typical star forming galaxies. }
    \label{fig:KS_comp}
\end{figure}

In the context of \squigglecomma, the high molecular gas fractions found in $\sim$half our sources is surprising given that these galaxies are, by all rest-optical tracers, quiescent \citep{Suess2017, Suess2022a, Bezanson2022a, Zhu2025}. To  illustrate this, in Figure \ref{fig:KS_comp}, we plot the star formation rate versus the molecular gas mass, with literature scaling relations illustrated as contours \citep{Saintonge2011, Tacconi2018, Freundlich2019}. The star formation rates of \squiggle galaxies measured in \cite{Suess2022a} place the full sample below this relation \citep[black points, see also][]{Bezanson2022a}. However, in contrast, literature post-starburst samples at similar redshift \citep[][shown as green diamonds]{Belli2021, Woodrum2022, Wu2023} all appear to have star formation rates that are largely within the scatter on star forming systems with similar $M_{H_2}$.

A key difference may lie in the way the star formation rates were inferred in these samples. All the aforementioned works measured their star formation rates using SED fits that incorporated rest-MIR SED information via Spitzer/MIPS 24 $\mu$m photometry, enforcing that the IR luminosity from attenuated stellar populations could produce that observation. In contrast, the \cite{Suess2022a} fits only utilized data from the rest-optical SED out to rest-frame $\sim2$ $\mu$m, and the fits to the \squiggle sample presented in that work were shown to under-predict WISE W3 and W4 photometry in sources that were detected. As such, it is possible that much of the tension between the inferred rest-optical star formation rates and the high molecular gas mass can be resolved by accounting for buried star formation that does not contribute in the rest-optical but produces significant IR luminosity. In the following section, we update the SED fits for the full \squigglecomma-CO(2-1) sample, addressing the IR SED in addition to a number of other modifications to test whether the previous fits were under-estimating the star formation rate of \squiggle post-starburst galaxies.

\subsection{Updated panchromatic spectrophotometric fitting} \label{subsec:SFH}

The SED fits presented in \cite{Suess2022a}, using \texttt{Prospector} \citep{Johnson2017,Leja2017, Johnson2021}, did not attempt to fit redward of W2, and therefore remained agnostic to the information contained at mid- and far-IR wavelengths. \texttt{Prospector} handles the dust SED by assuming energy balance, where the luminosity that is attenuated by dust is re-radiated in the mid- and far-IR with an SED that is described by the \cite{Draine2007} templates. The \cite{Suess2022a} fits acknowledged that the mid-IR SED of WISE-detected \squiggle galaxies could not be produced by the a set of fixed \cite{Draine2007} parameters they chose. However, the choice to fix these parameters, and a number of other choices related to the treatment of dust in the fitting, make the \cite{Suess2022a} fits ill-suited for answering questions about the presence of buried star formation, which has been suggested as a possible explanation for the gas-rich nature of many post-starburst galaxies \citep[e.g.,][]{Baron2023}.

For example, the \cite{Suess2022a} fits fixed the PAH-mass-fraction, $q_{PAH}$, to 2\%, limiting the ability of over-luminous PAHs that have been seen in post-starburst systems \citep[see][]{Smercina2018} to boost the mid-IR by shifting energy from cold dust to PAHs. The \cite{Suess2022a} fits similarly fixed the $U_{min}$ and $\gamma$ parameters that govern the shape of the dust SED to values typical of very cold dust, leaving no freedom for deviation from this distribution toward hotter dust. Additionally, the \cite{Suess2022a} fits did not include a mid-IR AGN torus, which can be included in \texttt{Prospector} using the \cite{Nenkova2008} torus templates \citep{Leja2018}, with normalization that is independent of the energy balance of the SEDs, providing a potential boost to the mid-IR flux without requiring any additional source of attenuated luminosity. 

\begin{deluxetable*}{ccccccccc}
\tabletypesize{\scriptsize}
\tablecaption{Physical properties of \squiggle galaxies. Only the three galaxies included in Figure \ref{fig:demo_IR_bright}, \ref{fig:demo_IR_medium}, and \ref{fig:demo_IR_faint} are presented, but the full table is available in a machine readable form online. \label{tbl:SED_fit}}
\tablehead{\colhead{ID} & \colhead{$ \log(M_\star / M_\odot) $} & \colhead{$ \mathrm{SFR} $\tablenotemark{a}} & \colhead{$ \mathrm{log(sSFR \ [yr^{-1}])}$} & \colhead{$ \tau_{BC} $\tablenotemark{b}} & \colhead{$ \tau_{ISM} $\tablenotemark{b}} & \colhead{$ \mathrm{Dust\ Index} $} & \colhead{$ t_\mathrm{PSB} $\tablenotemark{c}} & \colhead{$ L_{IR} $\tablenotemark{d}} \\
\colhead{} & \colhead{} & \colhead{[$M_\odot\ \mathrm{yr}^{-1}$]} & \colhead{} & \colhead{} & \colhead{} & \colhead{} & \colhead{[Gyr]} & \colhead{[$10^{11} L_\odot$]}
}
\startdata
\textbf{No Buried Star Formation:} & & & & & & & \\
J1157+0132 & $11.41^{+0.04}_{-0.04}$ & $17.57^{+9.55}_{-6.32}$ & $-10.16^{+0.19}_{-0.19}$ & $0.59^{+0.26}_{-0.20}$ & $0.55^{+0.08}_{-0.10}$ & $-0.05^{+0.11}_{-0.12}$ & $0.04^{+0.03}_{-0.01}$ & $6.25^{+0.96}_{-0.90}$ \\
J0910+0218 & $11.13^{+0.06}_{-0.08}$ & $14.86^{+6.34}_{-3.46}$ & $-9.97^{+0.22}_{-0.13}$ & $0.80^{+0.18}_{-0.13}$ & $0.82^{+0.09}_{-0.07}$ & $-0.55^{+0.09}_{-0.08}$ & $0.02^{+0.00}_{-0.00}$ & $5.86^{+1.20}_{-0.72}$ \\
J1141-0109 & $11.40^{+0.02}_{-0.03}$ & $0.00^{+0.08}_{-0.00}$ & $-14.89^{+2.38}_{-1.96}$ & $0.23^{+0.10}_{-0.12}$ & $0.19^{+0.07}_{-0.05}$ & $-0.45^{+0.25}_{-0.22}$ & $0.23^{+0.05}_{-0.03}$ & $0.64^{+0.18}_{-0.16}$ \\
... & ... & ... & ... & ... & ... & ... & ... & ... \\
\textbf{Buried Star Formation Allowed:} & & & &  & & & \\
J1157+0132 & $11.40^{+0.04}_{-0.05}$ & $75.68^{+25.60}_{-24.17}$ & $-9.52^{+0.13}_{-0.17}$ & $2.06^{+0.41}_{-0.64}$ & $0.48^{+0.09}_{-0.14}$ & $-0.07^{+0.08}_{-0.17}$ & $0.02^{+0.00}_{-0.01}$ & $8.67^{+1.44}_{-1.23}$ \\
J0910+0218 & $11.12^{+0.06}_{-0.05}$ & $26.58^{+15.50}_{-12.31}$ & $-9.71^{+0.22}_{-0.29}$ & $1.26^{+0.41}_{-0.48}$ & $0.81^{+0.08}_{-0.07}$ & $-0.56^{+0.10}_{-0.07}$ & $0.02^{+0.02}_{-0.00}$ & $6.47^{+1.25}_{-1.05}$ \\
J1141-0109 & $11.44^{+0.03}_{-0.05}$ & $0.00^{+0.13}_{-0.00}$ & $-14.47^{+2.17}_{-1.66}$ & $1.22^{+1.06}_{-0.74}$ & $0.21^{+0.06}_{-0.05}$ & $-0.26^{+0.19}_{-0.19}$ & $0.25^{+0.06}_{-0.04}$ & $0.61^{+0.14}_{-0.14}$ \\
... & ... & ... & ... & ... & ... & ... & ... & ... \\
\enddata
\tablenotetext{a}{Measured in the 10 Myr before observation}
\tablenotetext{b}{\texttt{dust1} and \texttt{dust2} in \texttt{fsps}.}
\tablenotetext{c}{Defined as the time at which the galaxy formed 99\% of the total stellar mass in the Gyr before observation.}
\tablenotetext{d}{Defined as the total luminosity of all dust components in our modeling.}

\end{deluxetable*}

Finally, the \cite{Suess2022a} fits adopt a two-screen \cite{Charlot2000} dust model, where all stars see dust with optical depth $\tau_{ISM}$, and young ($t<10^{7}$ yr) stars see an additional ``birth cloud" dust with optical depth $\tau_{BC}$. However, motivated by \cite{Wild2016} and by observations that show a relation between stellar- and Balmer decrement- $A_V$ \citep{Price2014}, tracing these two quantities, the \cite{Suess2022a} fits fixed $\tau_{BC}=\tau_{ISM}$. This assumption is quite conservative relative to other SED codes where the two parameters are left independent of one another \cite[for example, \texttt{MAGPHYS} allows the terms to vary independently and places a fairly weak prior on the total attenuation, see][]{daCunha2008}. Because the \squiggle post-starburst galaxies are by definition selected to look like low-dust A-star populations that lack emission, this prior effectively did not allow for any star formation to be buried behind optically thick dust.

Given the previous SED fitting, it is not actually clear on a per-galaxy basis: 1) if any additional star formation is necessary to explain the mid-IR luminosity, given that young stellar populations subjected to modest dust attenuation can still produce significant IR luminosity \citep[e.g.,][]{Wild2025}, and, 2) how much star formation the mid-IR detections (or upper limits) can tolerate given that there \textit{must} be significant IR contribution from dust heated by attenuated A stars. Here, we address those questions head on by fully folding our full set of constraints on the IR spectral energy distribution into our analysis and re-fitting the \squigglecomma-CO sample with IR dust templates under the assumption of energy balance. We make the following key modifications from the \texttt{Prospector} fitting procedure outlined in \cite{Suess2022a}:

\begin{enumerate}
    \item We include WISE W3 and W4 photometry presented (but excluded from the fits) in \cite{Suess2022a}. We exclude SDSS u-band imaging, as most of our sources are at the edge of the SDSS detection limit ($u_{AB}\gtrsim22$), and we have found that the uncertainties on supposed detections appear to be underestimated when inspecting these images. The exclusion of this photometry does not affect our fits. In total, our fits include the SDSS $griz$ photometry and WISE W1/W2/W3/W4 photometry, spanning $\lambda_\mathrm{rest}=0.3-13$ $\mu$m.  Additionally, for the 3 galaxies with photometry in the Herschel/SPIRE point source catalog at 250 $\mu$m ($\lambda_{rest}\sim150$ $\mu$m), we include those measurements in our fits. 
    \item The \cite{Draine2007} parameters that govern the reprocessed attenuated luminosity are left free, with $q_{PAH}$ (the PAH mass fraction) $\in [0.1,10]$, and $\gamma_e$ and $U_{min}$ (which, broadly speaking, control the temperature of the dust SED) $\in [0.0001,1]$ and $[0.1,25]$ respectively. We sample the PAH mass fraction linearly \citep[favoring the high PAH luminosity contribution seen in local post-starburst galaxies, see][]{Smercina2018}, and sample in log space for $U_{min}$ and $\gamma_e$ to allow for maximally flexible dust SEDs (as the dust SED changes logarithmically with these parameters). This model is scaled to the total attenuated luminosity for a given star formation history/dust law draw during the fitting.
    \item The \cite{Nenkova2008} AGN torus templates are also included in the fit, with a free AGN luminosity (parameterized as a fraction of the total bolometric luminosity of the galaxy) $\log(f_\mathrm{AGN})\in [-3,0.48]$ and $\log(\tau)\in[0.7,2.2]$ following \cite{Leja2018}. The luminosity of this AGN torus is not tied to any observed properties of the galaxy (e.g., rest-optical AGN indicators), although we note that as many as 20\% of the youngest \squiggle galaxies may host an AGN, as traced by the [OIII]/H$\beta$ line ratio \citep{Greene2020}.
    \item We modify the \cite{Suess2022b} post-starburst star formation history, adding one final bin that covers exactly the most recent 10 Myr of star formation. We do so because the two-screen \cite{Charlot2000} dust model assumed in \texttt{fsps} by default allows young ($t<10$ Myr) stars to see an additional ``birth cloud" dust screen that is added to the ``ISM" dust screen \citep{Conroy2009}. Forcing this bin of star formation to be independent from the rest of the star formation history allows dust-obscured star formation to be independent from the rest of the star formation in the final bin. The total length of time of the final two bins of star formation, $t_{last}$, is fit as a free parameter drawn from a flat (in log-space) distribution with a minimum time of 20 Myr and a maximum time of 15\% of the age of the universe at the redshift of observation. The final two bins have lengths of 10 Myr and $t_{last} - 10$ Myr, respectively.
    \item Finally, a number of priors from the \cite{Suess2022a} fits are relaxed. First, we do not include a prior on the star formation history based on the average star formation history of a mass- and redshift-matched galaxy in the Universe Machine \citep{Behroozi2019}; instead, we place a Student's t prior on the logarithmic ratio of star formation rate between neighboring bins (with $\sigma=0.3$ and $\nu=1$ for all but the final two bins, which have $\sigma=0.5$ to allow for a more rapid rise or fall). We also relax the mass-metallicity relation prior \citep[based on the][]{Gallazzi2005} relation that was used in \cite{Suess2022a}, allowing the metallicity to vary freely with $\log(Z/Z_\odot)\in [-0.4, 0.19]$. 
\end{enumerate}

\begin{figure*}
    \includegraphics[width=\textwidth]{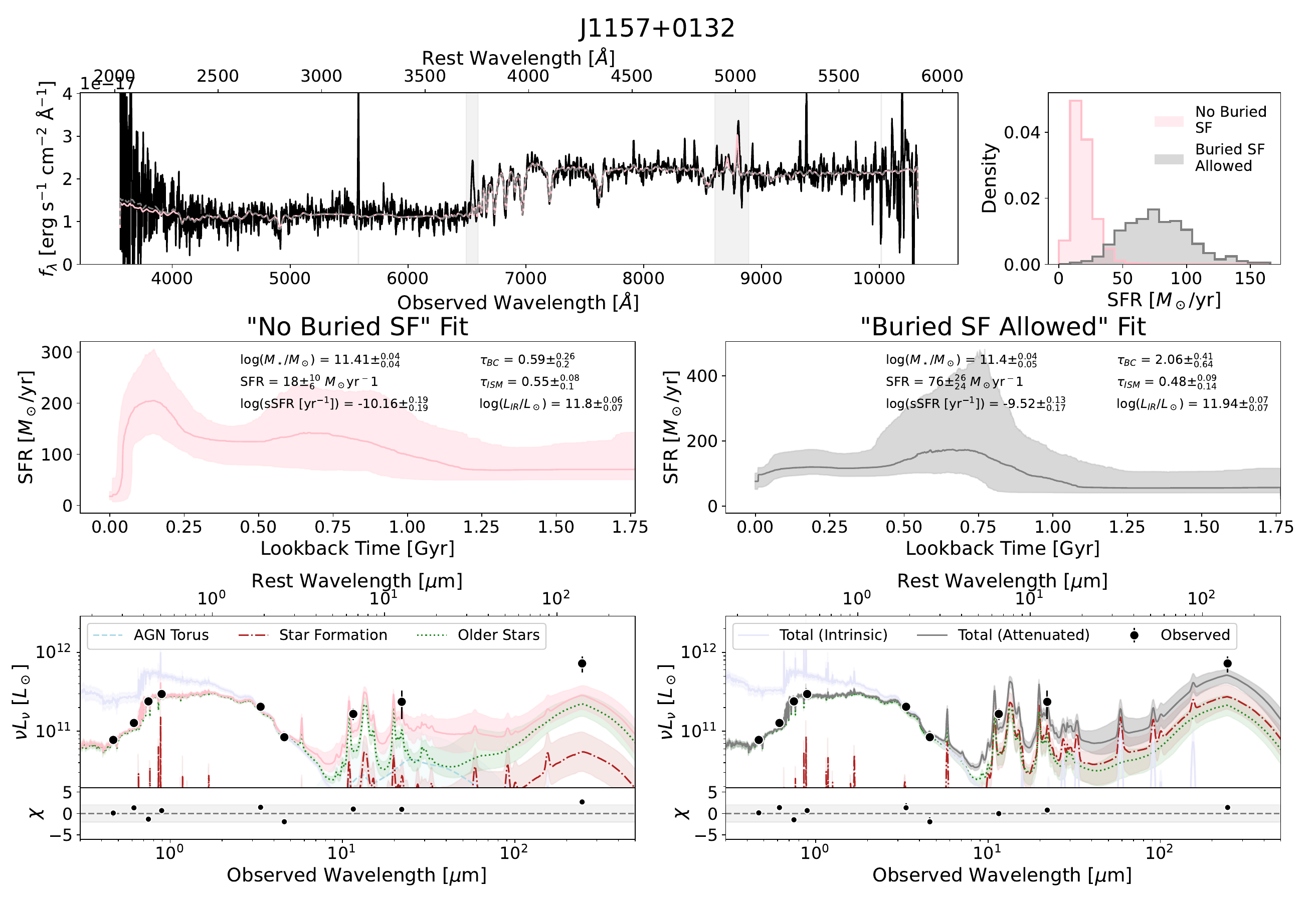}
    \caption{A demonstration of the new SED fitting adopted in this work, including mid- and far-IR photometry along with AGN and galaxy dust templates. Here, we highlight J1157+0132, a galaxy that is detected in the mid-IR, in ancillary Herschel/SPIRE 250 $\mu$m imaging, and is the most luminous in CO in our sample. In the top left, we show the SDSS spectrum (smoothed with a 5 pixel boxcar), along with the best fitting models for both the No Buried SF and Buried SF Allowed fits, which are virtually indistinguishable except for in [OII] and [OIII] emission (which is masked in the fit). In the top right, we show the posteriors for the instantaneous ($t<10$ Myr) star formation rates for these two fits. In the middle row, we show the best fitting star formation history (with 1$\sigma$ uncertainties) in the age period probed by our flexible bins, omitting the poorly constrained early universe fixed-bin star formation. In the bottom row, we show the best fitting SEDs, with individual components for the contributions of the AGN torus (light blue), star formation ($t<10$ Myr old stellar populations and the dust heated by them), and older stars ($t>10$ Myr stellar populations and the dust heated by them). We show the intrinsic stellar population (before dust attenuation) in magenta. We also show the residuals (data - model) of the photometric fit, with the shaded region indicating $\pm2$. In the No Buried SF fit essentially the entire IR SED is produced by the evolved, post-starburst population, but it is deficient by a factor of $\sim2$ relative to the 250 $\mu$m photometry. However, in the Buried SF Allowed fit, dust obscured star formation contributes a very similar fraction of the total IR luminosity to older stars, and the fit is better able to reproduce the observed Herschel photometry due to a small ($\sim0.15$ dex) boost to the total IR luminosity. \label{fig:demo_IR_bright}}
\end{figure*}

\begin{figure*}
    \includegraphics[width=\textwidth]{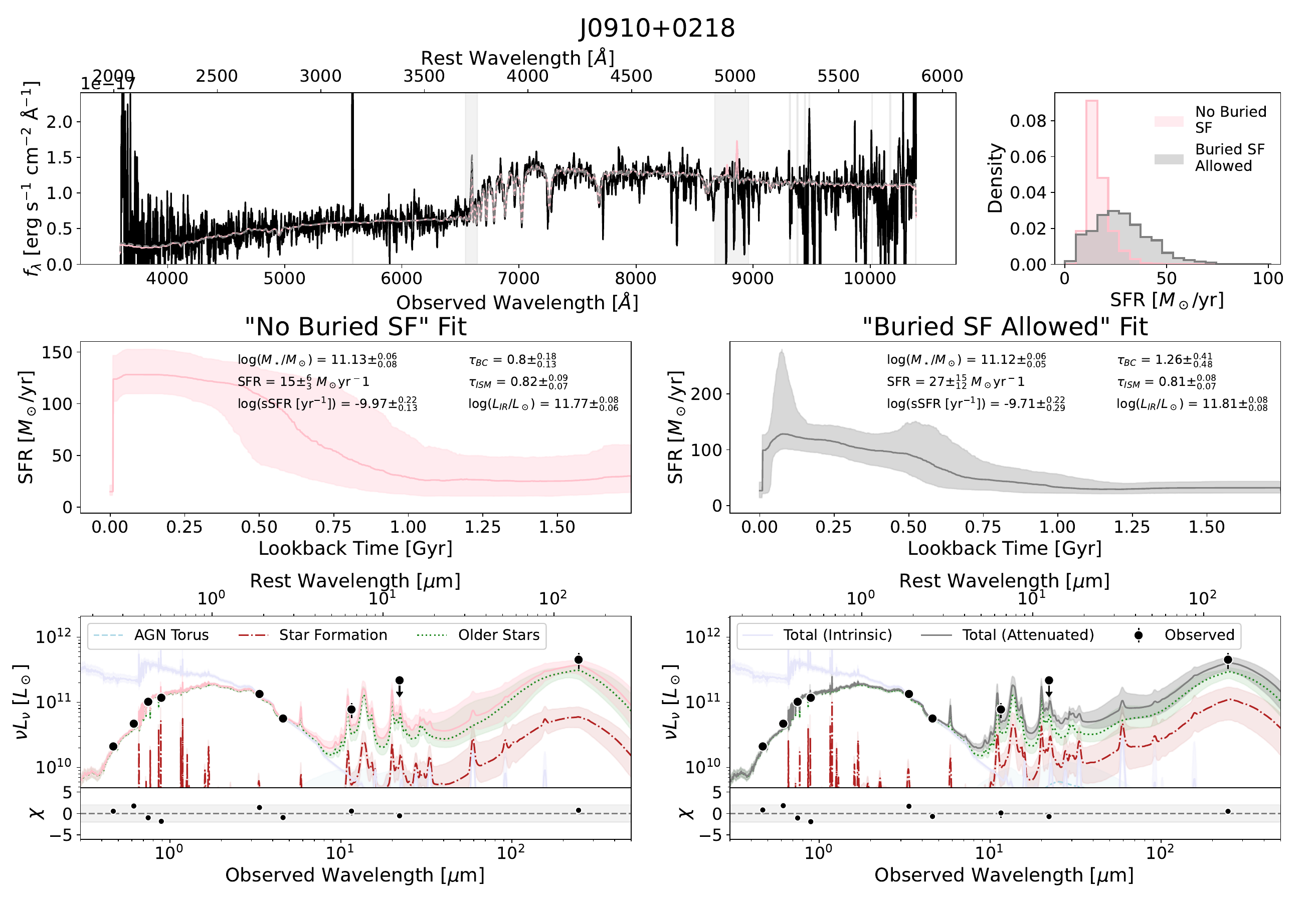}
    \caption{A demonstration of the new SED fitting adopted in this work, as in Figure \ref{fig:demo_IR_bright}. Here, we highlight J0910+0218, our second most CO-luminous galaxy that is detected in the mid-IR and in ancillary Herschel imaging. In contrast with J1157+0132 (see Figure \ref{fig:demo_IR_bright}), both the No Buried SF and Buried SF Allowed fits can produce provide a good match to the full SED, including Herschel. Even in the case where buried star formation allows for a somewhat higher star formation rate, the total IR luminosity is essentially identical between these models. \label{fig:demo_IR_medium}}
\end{figure*}

\begin{figure*}
    \includegraphics[width=\textwidth]{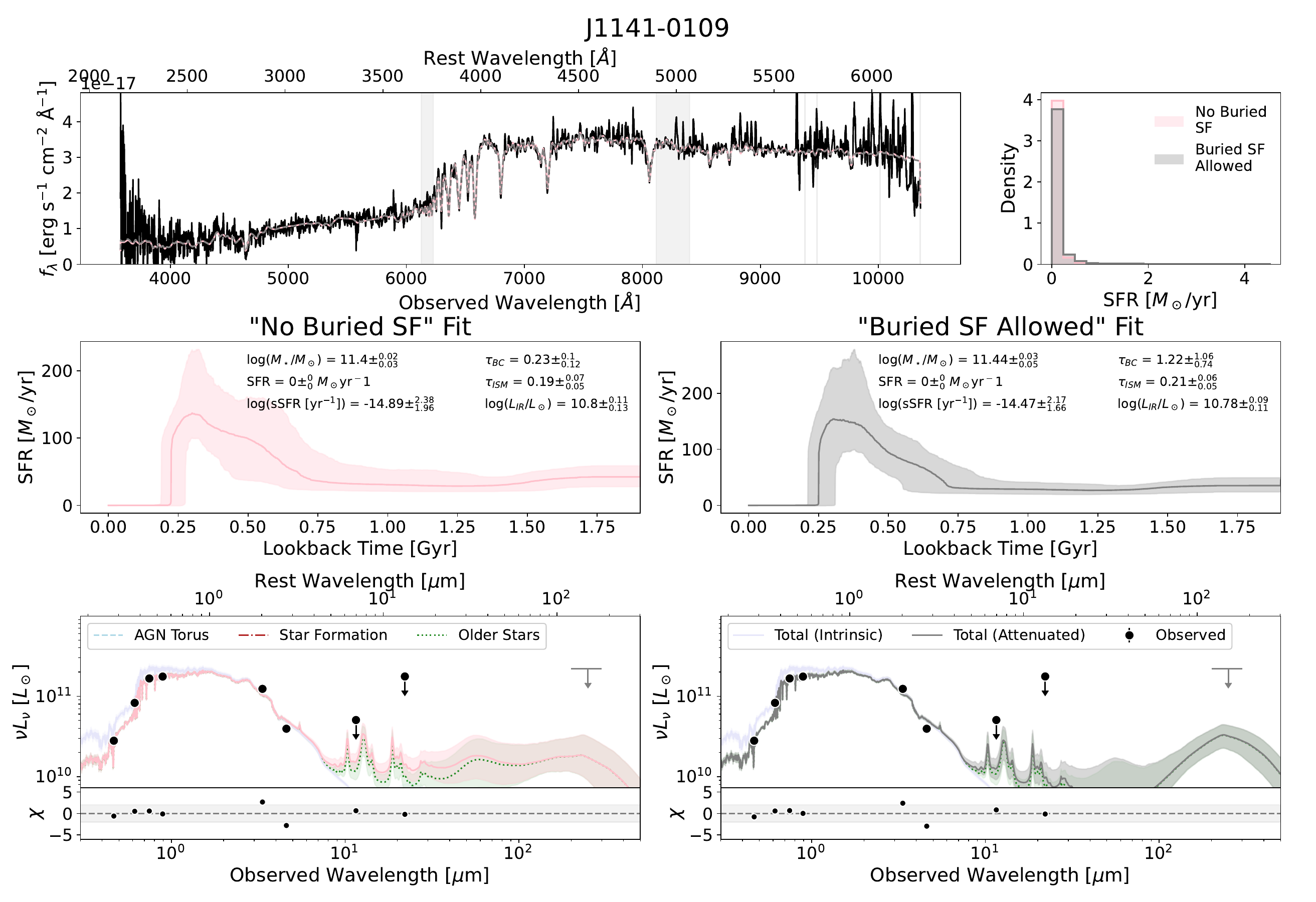}
    \caption{A demonstration of the new SED fitting adopted in this work, as in Figure \ref{fig:demo_IR_bright}. Here, we highlight J1141-0109, a galaxy that is not detected in the mid-IR, ancillary Herschel imaging, or in CO(2-1). While it is not included in the fit we also indicate the characteristic 3$\sigma$ uncertainty on a Herschel/SPIRE 250$\mu$m point source based on nearby objects in the point source catalog. In contrast with the IR detected galaxies, galaxies which are not IR-luminous return very similar star formation histories for both fits, regardless of the value of $\tau_{BC}$, because the mid-IR non-detections place a ceiling on the amount of star formation that can occur behind optically thick dust. \label{fig:demo_IR_faint}}
\end{figure*}

We perform two sets of fits, meant to test two assumptions about the relationship between the ISM dust and the birth cloud dust. In the first model, we mimic the \cite{Wild2020} \cite[motivated by][]{Price2014} prior choice that $\tau_{BC} \sim \tau_{ISM}$. However, in contrast with \cite{Suess2022a}, rather than simply fixing the parameters to have the same value, we instead place a Gaussian prior on the fraction between these two parameters, with $\mu=1$ and $\sigma=0.25$, leaving $\tau_{ISM}$ free $\in[0.0,2.5]$ with a flat prior. In a second set of fits, we allow for significantly more freedom in $\tau_{BC}$ to allow for buried star formation. To do so, we adopt a clipped Gaussian prior on $\tau_{BC}$, with $\mu=0.5$ and $\sigma=2.5$ \cite[though it is not the same functional form, this prior is very similar to the one adopted in MAGPHYS, see][]{daCunha2008}. The shape of the dust law for the ISM dust is parameterized with a dust index as in \cite{Kriek2013}, while the birth cloud dust law shape is fixed as a power law with a slope of -1. Throughout this work, we refer to these fits as ``No Buried SF" and ``Buried SF Allowed", respectively. 

Other than the aforementioned changes, our remaining modeling choices are largely similar to the \cite{Suess2022a} fits. We enforce a signal-to-noise ceiling of 20 on all photometry We use the \texttt{dynesty} dynamic nested sampling package \citep{Speagle2020}, the Flexible Stellar Population Synthesis (FSPS) stellar population synthesis models \citep{Conroy2009, Conroy2010}, the MILES spectral library \citep{Sanchez-Blazquez2006a,Falcon-Barroso2011}, and the MIST isochrones \citep{Choi2016, Dotter2016}. We assume a \cite{Chabrier2003} Initial Mass Function and fix the model redshift to the spectroscopic redshift as measured by SDSS. Before fitting, we convolve all models to the SDSS spectroscopic resolution, and we fit with an additional velocity dispersion term $\in [0,400]$ km/s. Nebular continuum and emission is turned on, and the nebular ionization parameter $\log(U)$ is left free $\in [-4,-1]$. However, we mask the [OII] and [OIII] lines as they are highly susceptible to ionization by shocks and AGN, both of which are not accounted for in our modeling framework. Finally, we utilize the \texttt{PolySpecModel} procedure with a 2nd order polynomial, and we include both a spectroscopic jitter term and the \texttt{Prospector} pixel outlier model. In Table \ref{tbl:SED_fit}, we show some of the derived parameters for two galaxies under both sets of fits; full tables of the entire \squiggle sample are available online.

\section{Results} \label{sec:results}

\subsection{Star formation histories}

In the previous section, we performed two sets of fits to the panchromatic SED and spectra of \squiggle post-starburst galaxies, one which used a commonly adopted prior where the ISM dust and the birth cloud dust are tied together (No Buried SF) and one which relaxed that prior to allow significantly more freedom in the birth cloud dust (Buried SF Allowed). Here, we explore the different conclusions about the star formation histories of post-starburst galaxies from the application of those priors.

In Figures \ref{fig:demo_IR_bright}, \ref{fig:demo_IR_medium}, and \ref{fig:demo_IR_faint}, we present example SED fits to three galaxies. In each figure, we first show the observed SDSS spectrum, with the best fitting model spectra for the No Buried SF (pink) and Buried SF Allowed (grey) models shown as dashed lines. In the bottom two rows, we show the median fitting (with 68\% confidence interval) star formation history (middle, showing only the 25\% of the galaxy's star formation history probed by the flexible bins and omitting earlier star formation for clarity) and SED (bottom, including photometric residuals normalized by the uncertainty) for each of the models. In the SED fit, we break down the spectrum into three components, active star formation (red) which shows the stellar and dust contributions from stellar populations with $t<10$ Myr, the contribution from older stars that formed $t>10$ Myr ago (green), and the mid-IR \cite{Nenkova2008} CLUMPY torus (blue, though in all of these sources, an AGN torus is not needed to fit the mid-IR). 
    
In Figure \ref{fig:demo_IR_bright}, we highlight J1157+0132, our most CO-luminous and Herschel-luminous source. For this source, the two sets of priors yield remarkably different star formation histories. The fit which does not favor buried star formation essentially reproduces the class of models presented in \cite{Suess2022a}, where the star formation histories of \squiggle galaxies indicate a large burst with SFR $\sim200 \ M_\odot \ \mathrm{yr^{-1}}$  and a sharp decline, in this case to a modest star formation rate (in the final 10 Myr bin) of $18\pm^{10}_{6}$ $M_\odot$/yr ($\log(\mathrm{sSFR \ [yr^{-1}]}) = -10.16 \pm ^{0.19}_{0.19}$). However, in contrast with the \cite{Suess2022a} fits that did not allow for any freedom in the shape of the FIR SED, our more flexible dust modeling has no problems producing a mid-IR SED that is in agreement with the data without the need for any additional buried star formation. The star formation rate constrained in this fit is quite similar to the original \cite{Suess2022a} fits, indicating that the problem producing the mid-IR photometry was not due to a lack of sufficient IR luminosity. Instead, it was simply the rigidity of the dust SED assumptions (for example, the \cite{Suess2022a} fits fixed $q_{PAH}=2$, whereas the fit we present here requires $q_{PAH} = 6.4\pm^{2.0}_{2.0}$) that drove the inability to produce the red IR continuum. However, while the predicted IR luminosity from the attenuation of the stars that formed in the $\sim100$ Myr old burst is nearly enough to account for all the mid-IR luminosity, this model undershoots the Herschel/PACs detection by a factor of 2.7, indicating that additional IR luminosity is needed to fully explain the FIR.

In our Buried SF Allowed modeling framework, this additional IR luminosity is provided by additional star formation behind optically thick dust, which limits its influence in the rest optical. The Buried SF Allowed fits can tolerate significantly higher star formation rates, with a highly uncertain best fitting SFR of $76\pm^{26}_{24}$ $M_\odot$/yr. In this class of solutions, the active star formation contributes comparably to the IR luminosity relative to the older stars, with the optically thick dust mirroring a sharp drop in star formation that can produce the deep Balmer absorption seen in the spectrum. Interestingly, despite a 0.4 dex discrepancy between the model and observed 250 $\mu$m flux in the No Buried SF model, the Buried SF Allowed model is able to reproduce the full SED with only a 0.14 dex shift in the total IR luminosity. This IR luminosity boost can be achieved even with star formation rates that are contained in the posterior of the No Buried SF framework, as evidenced by the overlapping posteriors. As such, while we conclude that in our modeling framework, buried star formation is required to produce this shift, the degree of buried star formation required is currently highly uncertain. J1157+0132 could indeed still be in the midst of its starburst, with optically thick dust attenuating its emission, but it could also be in a period of rapid decline from a recent period of much higher star formation rate during a $\sim100$ Myr old burst.

In Figure \ref{fig:demo_IR_medium}, we highlight J0910+0218, our second most CO-luminous and Herschel-luminous source. In contrast with J1157+0132, here both sets of fits do an equally good job at describing the mid-IR SED, as evidenced by the practically identical constraints on for the IR luminosity between the two models despite the allowance for a factor of $\sim2\times$ more star formation in the Buried SF Allowed fit. As such, in this source, which which is far more representative of our CO-detected sample than J1157+0132, we definitively cannot state that the data itself exhibits any strong preference for any level of buried star formation; the luminosity from the attenuated $\sim$100 Myr stellar populations that dominate the rest-optical continuum is more than enough to produce the observed FIR SED.

Finally, in Figure \ref{fig:demo_IR_faint}, we highlight J1141+0109, a galaxy which is not detected in any IR band or in CO(2-1). In contrast with J1157+0132 and J0910+0218, the two sets of fits to J1141-0109 return essentially identical star formation histories and physical properties. Because the galaxy is not detected in the infrared, the combination of the W3/W4 non-detections place a hard cap on the maximum IR luminosity that is tolerable in the fits. As such, there is very little room to add an appreciable contribution from buried star formation even when the priors are more permissive to hiding emission in the rest-optical. This is reflected in the overlapping posteriors for the star formation rate and the IR luminosity, even when the best fitting $\tau_{BC}$ is significantly greater than $\tau_{ISM}$.

\begin{figure}
    \centering
    \includegraphics[width=0.48\textwidth]{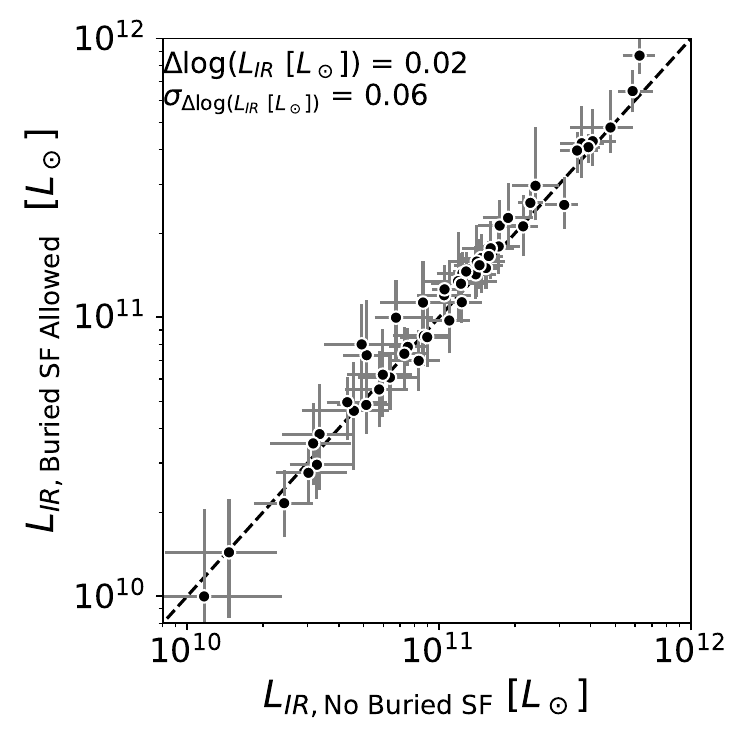}
\caption{Comparison of the inferred total IR luminosity (defined as the total luminosity coming from all dust components in our modeling) between the two sets of models. We label the median logarithmic offset and scatter in the offset (Buried SF Allowed - No Buried SF). The two fits recover extremely similar total IR luminosity, which is well constrained by the combination of WISE and 2mm data under the assumption of the \cite{Draine2007} galaxy dust templates and the CLUMPY torus templates \citep{Nenkova2008}. Any increase in the fractional contribution to the IR luminosity from buried star formation must be compensated for by a decrease in the fractional contribution from older stars, the mid-IR AGN luminosity, or changes to the dust temperature distribution or PAH mass fraction.} \label{fig:model_comparisons}
\end{figure}

In Figure \ref{fig:model_comparisons}, we compare the full set of galaxies in the inferred total IR luminosity (defined as the sum of the luminosity of all dust components included in our fits, including an AGN torus). The vast majority of the galaxies in \squiggle show essentially no deviation in their total IR luminosity between the two fits, indicating not only that producing the observed IR luminosity is possible with fits that do not require buried star formation, but that the attenuated luminosity of the moderately dusty A-type populations required to fit these post-starburst spectra is already so high that there is little room in the dust SED to hide a contribution from a starburst that is invisible in the rest-optical. As such, the median shift in $\log(L_{IR})$ from the No Buried SF fit to the Buried SF Allowed fit is only 0.02 dex, with a scatter that encompasses 0, and even in the most extreme cases, J1157+0132 (the most IR luminous galaxy in the sample), there is not room for more than a factor of $\sim2$ more IR luminosity than is implied in the fits without buried star formation.

This manifests in star formation rates that are boosted only moderately in the Buried SF Allowed fits, with a median shift of 0.5 dex in sources with a star formation rate $>1 M_\odot \ \mathrm{yr^{-1}}$ where constraints from this kind of fitting are meaningful \citep[see][]{Suess2022b}. This factor of $\sim3$ shift, even under the assumption that there is a maximal amount of star formation behind optically thick dust, is quite small given the freedom of the fits, and in all cases, the posterior for the star formation rate measured in the Buried SF Allowed fit overlaps significantly with posterior for the the star formation rate measured under the No Buried SF priors. 

\begin{figure*}
    \centering
    \includegraphics[width=0.49\textwidth]{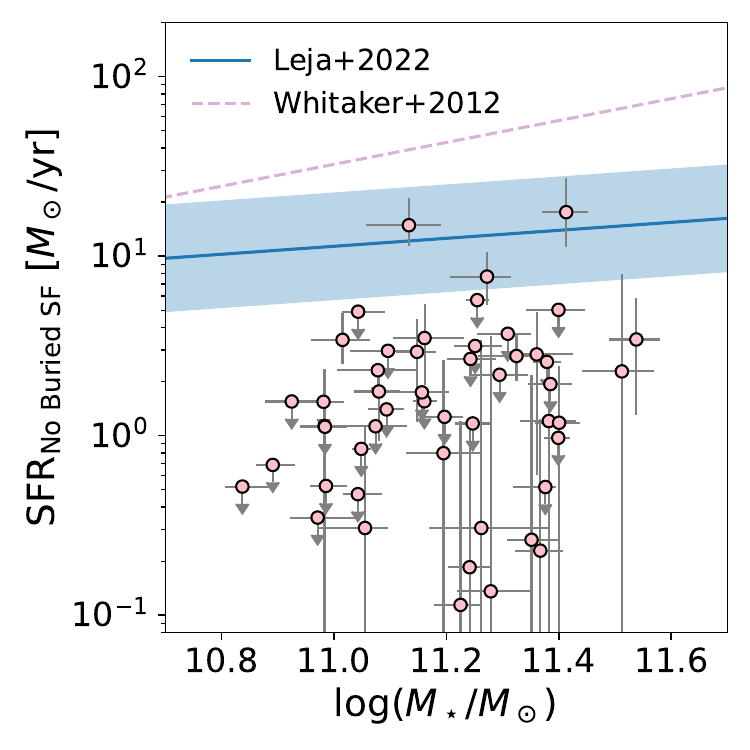}
    \includegraphics[width=0.49\textwidth]{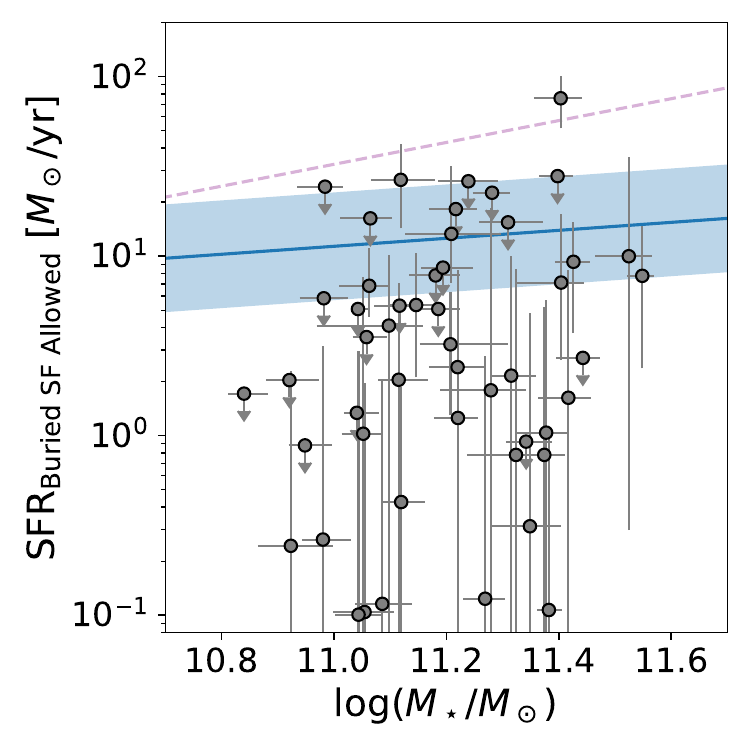}
    \caption{The star forming main sequence, with the best fit from \cite{Leja2022} at $z=0.7$ shown (with 0.3 dex scatter in blue) and the \cite{Whitaker2012b} fit shown in purple. On the left, we show the measurements from our No Buried SF fits (pink), and on the right, we show our Buried SF allowed fits. For clarity, for all star formation rates where the median is below 0.1 $M_\odot \ \mathrm{yr}^{-1}$, we show the 3$\sigma$ upper limit. Maximal buried star formation can raise the star formation rates of \squiggle galaxies, with a median offset of 0.5 dex in sources with $\mathrm{SFR_{Buried \ SF \ Allowed}}>1 \ M_\odot \ \mathrm{yr}^{-1}$, to place them at most on the \cite{Leja2022} main sequence \citep[but still essentially allow below the][main sequence]{Whitaker2012b}. However, there is considerable uncertainty in those dust obscured star formation rates, and in every model, the posterior for the star formation rate in the Buried SF Allowed fit overlaps significantly with the posterior of the No Buried SF fit.}
    \label{fig:SFMS}
\end{figure*}

We caution that the tight constraints on the IR luminosity in these fits, and subsequently, the star formation rate, in many ways rely on the validity of the assumption that the \cite{Draine2007} dust templates span the full range of possible dust temperature and grain size distributions such that our extrapolations from our rest-frame 10 $\mu$m (and, occasionally, a single point at $\sim$150 $\mu$m) observations are constraining. Dust that emits with a distribution that maximizes output at warm temperatures where we lack constraints could allow for more IR luminosity, and therefore more star formation. Future work, ideally with better FIR data from an observing facility like PRIMA \citep{Moullet2023} or with more robust dust-obscured star formation rate tracers, could better allow for these degeneracies to be broken. For now, we proceed forward under the assumption that our two sets of measured star formation histories span a reasonable range of the possible present day star formation activity in \squiggle post-starburst galaxies. In the next section, we investigate how the changes in the inferred star formation rate would affect the interpretation of where these galaxies lie relative to the star forming main sequence.

\subsection{The star forming main sequence}

Many works have found that many post-starburst galaxies lie systematically below the star forming main sequence and the SFR-$M_{H_2}$ relation for normal star forming galaxies for a variety of rest-optical SFR tracers \citep[e.g.,][]{French2015, Suess2017, Belli2021, Bezanson2022a}, though with considerable disagreement between different star formation rate tracers. 
\cite{Baron2023} argues that the high IR luminosities of post-starburst galaxies imply star formation rates that are well above the main sequence, resolving all tension related to the depletion time but requiring that significant star formation is occuring behind optically thick dust.
However, it has also been argued that direct conversions from the IR luminosity to star formation rate in these systems can significantly overestimate the true instantaneous star formation rate, as heating from more evolved stars (especially young A type stars formed on timescales of $\sim100$ Myr before observation) can contribute significantly to the IR luminosity \citep{Utomo2014,Hayward2014, Leja2019, Wild2025}. 
    
In the previous section, we showed that we can measure instantaneous (10 Myr) star formation rates under two sets of assumptions about the level of allowed buried star formation that simultaneously produce the UV and the IR for our galaxies. In Figure \ref{fig:SFMS}, we compare the star formation rates we infer under these two sets of assumptions against the star forming main sequence at $z=0.7$ from \cite{Leja2022}, with the No Buried SF fit shown in pink on the left and the Buried SF Allowed fit in grey on the right. For clarity, galaxies with a median SFR $<0.1$ $M_\odot$/yr are represented by their $3\sigma$ limits. 

As stated previously, the median shift in the measured SFR under a much more permissive prior for the dust around birth clouds is only $0.5$ dex. While this is enough to boost some galaxies onto the star forming main sequence (SFMS), especially when upper limits are considered, the vast majority of galaxies are still consistent with lying below this relation, and only one galaxy (J1157+0132, see Figure \ref{fig:demo_IR_bright}) is able to hide enough IR luminosity that it can be consistent with being in the midst of an active starburst. As such, we conclude that the \squiggle post-starburst galaxies can at most be hosting star formation rates that are consistent with lying on the \cite{Leja2022} main sequence, which is notably lower than other published sequences such as the \cite{Whitaker2012b} relation (also shown, in purple) at high masses due to the accounting for the heating from old stars in the conversion from IR luminosity to the star formation rate. 

Even with buried star formation allowed, the \squiggle galaxies still lie below the \cite{Whitaker2012b} relation, which assumes that the entirety of the IR luminosity is produced by active star formation. This suggests that while \squiggle galaxies may still be forming star stars at moderate levels, they are not still in the midst of active starbursts, and an appreciable part of their IR luminosity must come from evolved stellar populations. And, once again, the data provides no strong evidence for the higher star formation rates measured with more permissive priors; moderate star formation rates simply cannot be ruled out with existing data.

\newpage 

\subsection{The molecular gas content as a function of age}

\begin{figure*}
    \centering
    \includegraphics[width=0.49\textwidth]{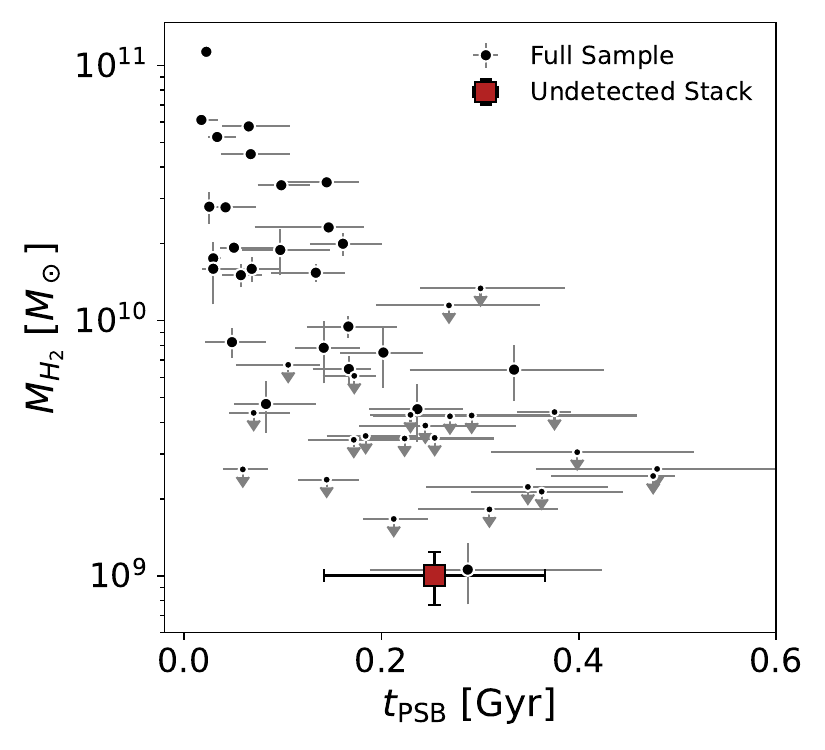}
    \includegraphics[width=0.49\textwidth]{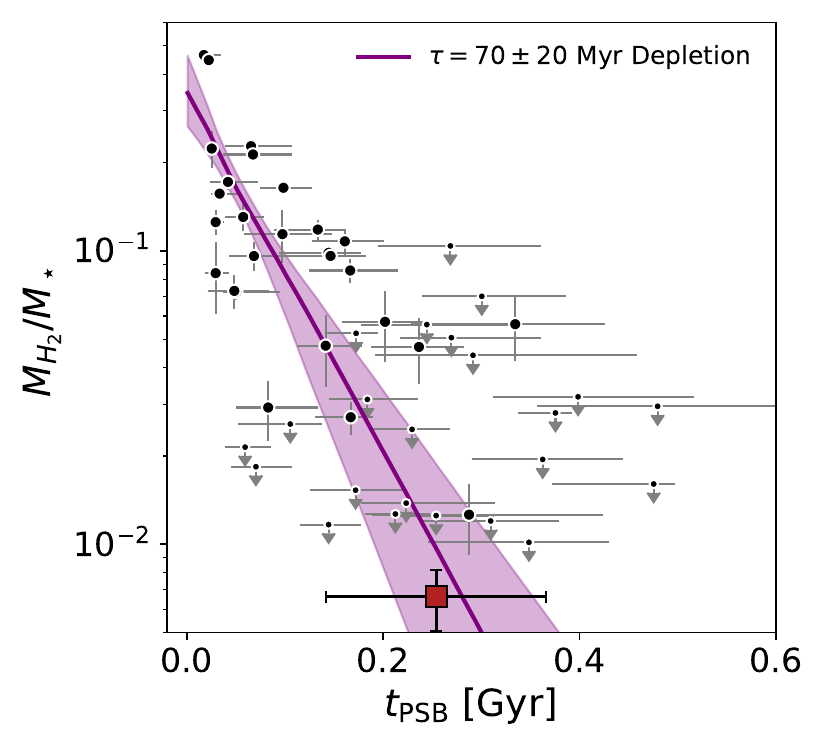}

    \caption{The $M_{H_2}$ mass (assuming $r_{21}$ and $\alpha_{CO}=4.0$, left) and the molecular gas fraction (right) versus the post-starburst time (defined as the time at which the galaxy formed 99\% of the total mass formed in the Gyr before observation, as measured from the Buried SF Allowed fits). We show \squiggle points in black, with CO-detected galaxies as large symbols and CO upper limits as smaller symbols. A stack of all non-detections is shown as a red square at the median $t_\mathrm{PSB}$. Similar to the trend with the time since quenching reported in \cite{Bezanson2022a}, we find that only the youngest \squiggle galaxies are CO-luminous. In purple, we show a best fitting exponential to the gas fraction as a function of age with a decay timescale of $70\pm20$ Myr, implying that galaxies must clear out their remaining molecular gas reservoirs in $\lesssim140$ Myr to match the observed trend.}
    \label{fig:tpsb_LCO}
\end{figure*}

Previous analysis of \squiggle CO(2-1) observations in \cite{Bezanson2022a} found a clear bi-modality in the observed properties of galaxies, with the youngest \citep[time since quenching $\lesssim$200 Myr, see][]{Suess2022a} exhibiting highly luminous CO and the older galaxies remaining undetected. Similar age trends are seen in \cite{French2018a} as a function of the post-burst age.

Here, we further investigate that trend with our full sample. Rather than using the time since quenching, which requires a sharp drop in measured star formation rate (which does not occur in all acceptable models under the Buried SF Allowed priors, see Figure \ref{fig:demo_IR_bright}), we instead define $t_\mathrm{PSB}$ as the lookback time at which a galaxy formed 99\% of its stars in the Gyr before observation. This quantity has the advantage of being largely insensitive to the choice of prior; the median scatter in $t_\mathrm{PSB}$ between the two sets of priors is 0, with a scatter of 0.1 dex. In Appendix \ref{sec:timescale_comp}, we show the comparison between this quantity and the ``time since quenching" measured in \cite{Suess2022a}, finding very good agreement. 

In Figure \ref{fig:tpsb_LCO}, we show the trend between the molecular gas mass and molecular gas fraction (assuming $r_{21}=1.0$ and $\alpha_{CO}=4.0$) and $t_\mathrm{PSB}$. We find that the bimodality presented in \cite{Bezanson2022a} holds for our larger and more representative \squiggle sample; essentially all CO(2-1) luminous galaxies have $t_\mathrm{PSB}<200$ Myr, and a stack of all non-detections is only detected at the 2$\sigma$ level with 1-2 dex lower CO(2-1) luminosity than our detected galaxies. We fit an exponential decay to the gas fraction as a function of age, estimating uncertainty by running 1000 fits to draws from the stellar mass and age posteriors for the ensemble of galaxies and incorporating the uncertainty in the gas mass in the gas fraction at each stage. We find a characteristic timescale of decay of $70\pm20$ Myr. In Appendix \ref{sec:timescale_comp}, we show that this qualitative trend is robust to different definitions of age, and the decay timescale is robust within a factor of 2. If the \squiggle sample is indeed tracing a single population drawn from a distribution of post-burst age, this implies galaxies must fully deplete their gas content within $\sim250$ Myr of quenching, and our detected galaxies must deplete their remaining fuel within $\sim140$ Myr (two e-folding times). In the following section, we explore the possible physical scenarios that might explain this trend.

\newpage

\section{Why does the CO(2-1) luminosity rapidly fade in the wake of a starburst?} \label{sec:tdep}

\begin{figure*}
    \centering
    \includegraphics[width=0.48\textwidth]{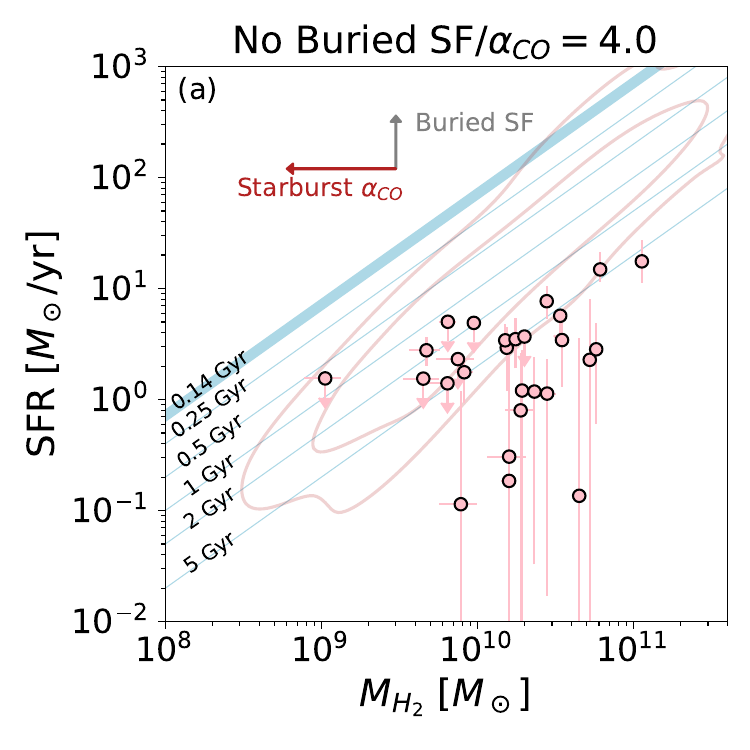}
    \includegraphics[width=0.48\textwidth]{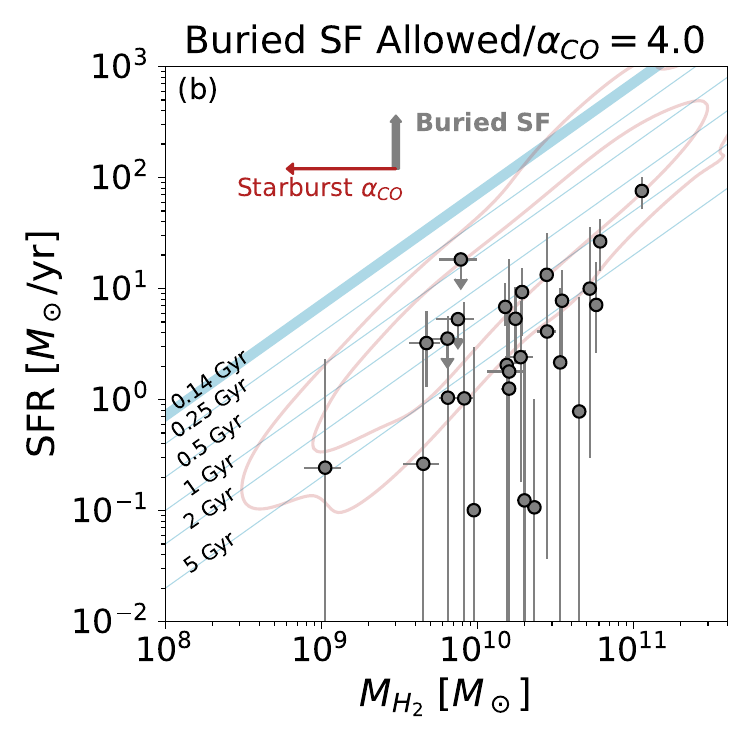}
    \includegraphics[width=0.48\textwidth]{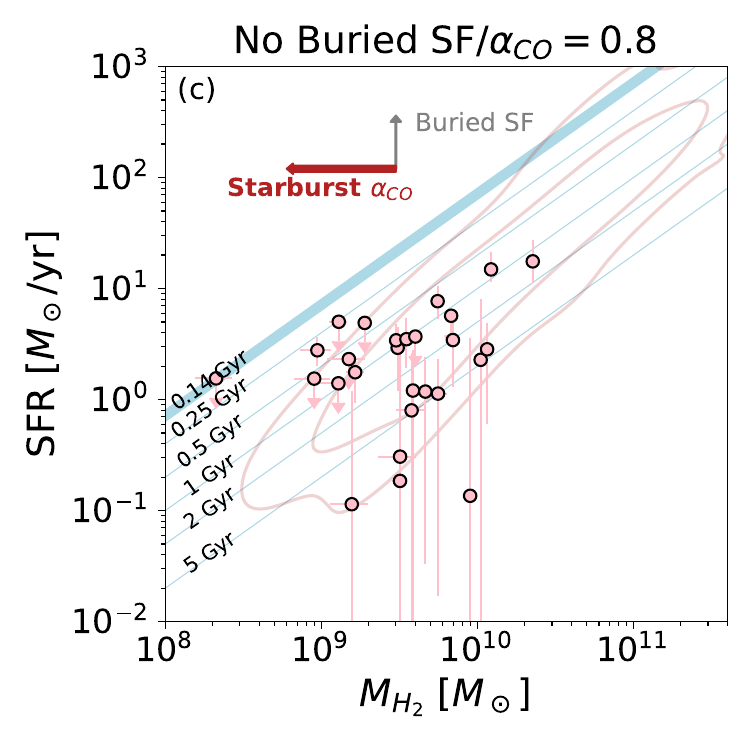}
\includegraphics[width=0.48\textwidth]{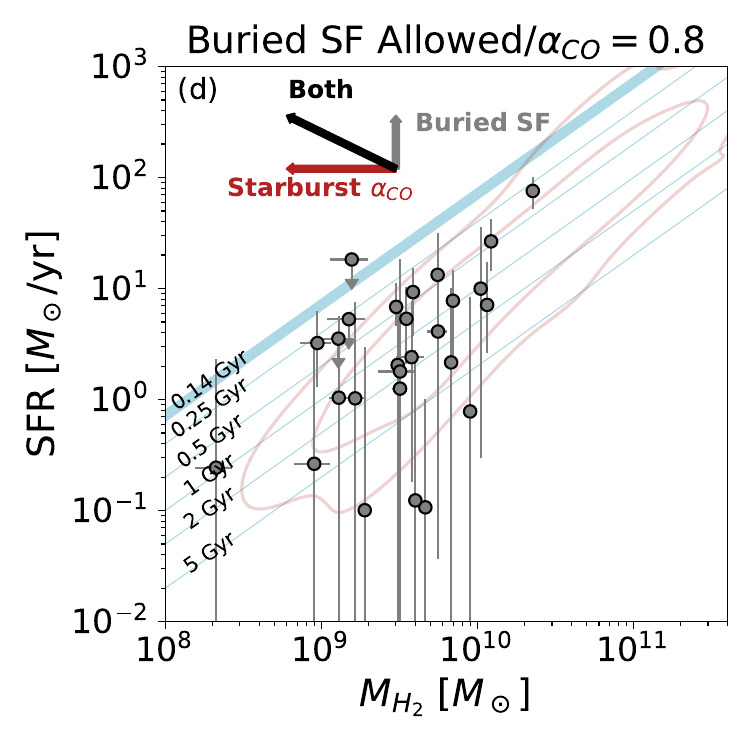}
    \caption{The resolved Kennicutt-Schmidt relation for the CO-detected \squiggle galaxies under four sets of assumptions: No Buried SF, $\alpha_{CO}=4.0$ (top left), No Buried SF, $\alpha_{CO}=0.8$ (bottom left), Buried SF Allowed, $\alpha_{CO}=4.0$ (top right), and Buried SF Allowed, $\alpha_{CO}=0.8$ (bottom right). In galaxies where the median star formation rate is less than 0.1 $M_\odot$/yr, we instead show the 3$\sigma$ upper limit. As background contours, we show all detected galaxies with $\log(M_\star/M_\odot)>10.8$ from the COLDGASS Survey \citep[$z\sim0$,][]{Saintonge2011} and PHIBSS/PHIBSS2 \citep[$z=0.5-2.5$,][]{Tacconi2018,Freundlich2019}, assuming $\alpha_{CO} = 4.0$.  In blue, we show lines of constant depletion time for the molecular gas, with the target depletion time of $\sim140$ Myr highlighted as a thick line. In the bottom right panel, we show vectors that indicate the typical shift that comes from changing one of these assumptions: 0.7 dex in $M_{H_2}$ (from  changing to a starburst $\alpha_{CO}$ assumption) and 0.5 dex in star formation rate. Even under the assumption that all the youngest ($t<10$ Myr) stars in every \squiggle are behind optically thick dust cannot place the galaxies in a part of SFR vs. $M_{H_2}$ space that results in depletion times that align with the trend in Figure \ref{fig:tpsb_LCO}. Similarly, a modification to $\alpha_{CO}$ can raise the depletion times significantly, but at the low inferred star formation rates, the majority of the galaxies still fall below the target depletion timescale of 140 Myr. Even the combination of these effects is not enough to resolve the tension with the observed depletion times.}
    \label{fig:tdep}
\end{figure*}

In the previous section, we demonstrated that the previously identified trend between CO luminosity and age wherein essentially all CO-luminous post-starburst galaxies have ages $\lesssim200$ Myr \citep{French2018a, Bezanson2022a} holds for the full \squiggle sample. This, along with the low luminosity of stacked non-detections, indicates that if the trend in Figure \ref{fig:tpsb_LCO} is taken at face value and these galaxies represent an evolutionary sequence, that the CO luminosity must drop by an order of magnitude on a timescale of $\lesssim140$ Myr. 

Here, we confront this trend by asking whether the depletion times of \squiggle galaxies (the time it would take to remove their gas reservoirs if they maintain their current star formation rate) can be commensurate with this implied $\sim140$ Myr timescale. Essentially all studies of the molecular gas content in post-starburst galaxies bake two key assumptions into their analysis \citep[e.g.,][]{French2015, Suess2017, Smercina2018, French2018a, Belli2021, Bezanson2022a, Woodrum2022, Zanella2023}:

\begin{enumerate}
    \item There is no significant buried star formation, and the low attenuation measured from stellar emission can be extrapolated using a relation similar to \cite{Price2014} to infer the attenuation around birth clouds.
    \item The Milky Way CO-to-$H_2$ conversion holds in these systems.
\end{enumerate}

Using the resolved Kennicutt-Schmidt relation \cite[SFR vs $M_{H_2}$,][]{Schmidt1959, Kennicutt1998}, in this section we systematically test how varying these assumptions effects the measured depletion time ($M_{H_2}$/SFR) of our systems, which should be similar to 100 Myr if \squiggle galaxies can passively evolve along the $L_{CO(2-1)}$ versus age relation. In Section \ref{subsec:standard}, we first present our fiducial model under the standard assumptions of no buried star formation and a Milky Way $\alpha_{CO}$. In Section \ref{subsec:buried}, we explore the effect of buried star formation by using the inferred star formation rates from the Buried SF Allowed fits. In Section \ref{subsec:lower_mass}, we explore the effect of relaxing the assumption of a Milky Way $\alpha_{CO}$, and also discuss the potential effects of merger-driven tidal stripping and/or outflows. In Section \ref{subsec:data_support}, we discuss whether there is evidence in our data or in the literature to support either of these modifications to the standard assumptions about the state of the post-starburst ISM. Finally, in Section \ref{subsec:rejuvenation}, we discuss the possibility that the trend in $L_{CO(2-1)}$ versus $M_{H_2}$ is not a trend at all and that \squiggle post-starburst galaxies may rejuvenate before re-entering the quiescent sequence.

\subsection{Can \squiggle galaxies deplete their gas under standard assumptions?} \label{subsec:standard}

We begin by exploring the standard assumptions of a Milky Way-like $\alpha_{CO}$ and star formation rates that are inferred from the rest optical under the assumption that there is no additional component behind optically thick dust. In the top left panel of Figure \ref{fig:tdep}, we show the \squiggle sample under the typical No Buried Star Formation and $\alpha_{CO}=4$ assumptions, replicating the finding from \cite{Bezanson2022a} that these galaxies lie below the relation with depletion times that are much larger than the characteristic CO-fading timescale and that post-starburst galaxies are offset from the population of typical star forming galaxies \citep{Saintonge2011, Tacconi2018, Freundlich2019}. It is not trivial that the entire sample still lies below the resolved Kennicutt Schmidt relation in the current fits, since unlike \cite{Suess2022a}, we explicitly include the mid- and far- infrared fluxes that are luminous in some of sources. In other words, there is sufficient heating from A-type stars to satisfy the long wavelength constraints in our No Buried Star Formation model without adding significant star formation under the assumption that $\tau_{ISM}\sim\tau_{BC}$, though we cannot rule out buried star formation from this fact alone.

Given that we recover the same location in the Keniccut Schmidt relation for \squiggle galaxies as \cite{Bezanson2022a}, we also reach the same conclusion with regard to their depletion times. Under standard assumptions, we measure depletion times $>1$ Gyr, with the bulk of the population lying closer to $\sim5$ Gyr, far longer than the $\sim140$ Myr timescale necessary to explain the falling CO luminosity. Therefore, if \squiggle galaxies do indeed evolve from CO luminous to CO faint on the observed timescale, there must be some modification to either their inferred star formation rate or the total gas mass consumption that is necessary in that time period. In the following sections, we explore both those possibilities.

\subsection{Buried star formation?} \label{subsec:buried}

We now confront the possibility of deeply buried star formation in our systems. As we demonstrated in Section \ref{sec:results}, there is some latitude to raise the star formation rates of \squiggle post-starburst galaxies by relaxing assumptions about the relationship between the ambient ISM dust and the birth cloud dust, with the potential to raise a typical galaxy's star formation rate by $\sim0.5$ dex. We stress again that this preference is not strongly favored by the data, as the vast majority of our galaxies are not detected in the mid-IR, and even those that are detected in W3/W4 can be well-modeled under the No Buried SF priors. Additionally, studies of low-z post-starburst galaxies have found good agreement between observed $H\alpha$ star formation rates and IR tracers like \citep{Smercina2018, Luo2022, French2023a}, providing evidence against dust obscured star formation beyond what is expected from standard prescriptions for the dust geometry. However, we cannot rule out the possibility that this buried star formation truly is there, resulting in a significant reduction in the depletion time.

In Figure \ref{fig:tdep}b, we again show the star formation rate versus the molecular gas mass, this time using the star formation rate measured from the Buried SF Allowed fits (and indicating the typical shift from the previous panel with a bold grey vector). Perhaps predictably, a factor of $\sim3$ boost in the typical star formation rate is not enough to resolve the order-of-magnitude tension between the observed and predicted CO depletion times, and the majority of the sources are consistent with depletion times closer to $\sim1$ Gyr than $\sim140$ Myr, especially in the most CO-luminous sources. 

Therefore, we conclude that even if there is significant star formation occurring behind optically thick dust in our galaxies, the implied star formation still could not deplete the gas quickly enough. Depletion of $\sim$a few $\times 10^{10}$ $M_\odot$ of molecular gas in $\sim100$ Myr requires star formation rates $\gtrsim100$ $M_\odot$/yr. Such high star formation rates cannot coexist with the strong Balmer absorption, the lack of strong emission lines, and the mid- and far-IR limits on the total IR luminosity within our modeling framework.

\subsection{Systematically lower $H_2$ masses?} \label{subsec:lower_mass}

After testing the potential impact of buried star formation, we now turn to the other lever arm in our depletion time calculation--the total $H_2$ mass that must form into stars to explain the rapidly dropping CO luminosity. Many works have shown a preference for mergers among post-starburst samples \citep[e.g.,][]{Alatalo2016b, Sazonova2021, Wilkinson2022,Ellison2024_merger}, and that same preference manifests in a very high merger fraction within \squiggle \citep[$\sim70\%$ in our youngest galaxies, see][]{Verrico2023}. Gas rich mergers have the potential to systematically lower the CO-to-$H_2$ conversion by allowing more CO luminosity per-unit-$H_2$ to escape a turbulent, more diffuse post-merger ISM \citep[e.g.,][]{Narayanan2011, Bournaud2015, Renaud2019}, and we investigate the impact of the assumed $\alpha_{CO}$ on our conclusions. In order to do so, we adopt a ULIRG-like $\alpha_{CO}=0.8$ \citep{Bolatto2013}. In Figure \ref{fig:tdep}c and \ref{fig:tdep}d, we illustrate the effect of this shift CO-detected \squiggle galaxies, with No Buried SF models on the left and Buried SF Allowed models on the right.

A 0.7 dex shift in the inferred $H_2$ mass from an altered $\alpha_{CO}$ is not enough to place the majority of the galaxies at depletion times where the $L_{CO(2-1)}$ versus age trend can be fully explained by the low-lying star formation in the No Buried SF models. Even combining this effect with buried star formation cannot resolve the tension; while many sources can reach depletion times of $\sim250$ Myr, the observed $\sim140$ Myr target remains elusive as none of the \squiggle galaxies are consistent with still being in the midst of active starbursts. 

We also address one other possibility related to lowering the $H_2$ mass in these post-starburst galaxies: the mechanical removal of star forming material via outflows and mergers. Previous works leveraging higher resolution CO(2-1) observations have shown that as much as half the CO luminosity in luminous \squiggle sources lies in tidal features on scales of tens of kiloparsecs \citep[][]{Spilker2022, Donofrio2025}, as well as in neighboring galaxies that may have recently interacted with \squiggle systems (Kumar et al. 2025 submitted). Even in the low-resolution data we present in this work, similar extended CO can be seen most clearly in J1157+0132 (see Figure \ref{fig:most_gas_rich}), our most CO-luminous system, where extended CO along the tidal features is kinematically distinct from the core of the galaxy. We also note that such extended CO luminosity is seen in co-eval massive starbursts and may be partially driven by outflows \citep{Geach2018}; similar evidence for outflows playing a large role in ISM removal has been found in high-redshift quiescent systems \citep[][]{Belli2023, Sun2025_PSBoutflow}, which could potentially drive the extended CO seen in the aforementioned works. However, in the absence of higher resolution CO maps, we can only speculate, and at least at present, the CO centroid and velocity are well-aligned with the rest optical locations of our galaxies, suggesting that the majority of the molecular gas in \squiggle post-starbursts will not be fully removed by any present outflow.

\subsection{Do the data support modifications to the star formation rate or molecular gas content?} \label{subsec:data_support}

In the previous sections, we demonstrated that while even the combination of buried star formation and a and a reduction in the CO-to-$H_2$ conversion cannot totally explain the trend of post-starburst age versus gas fraction, the depletion times can be lowered by an order of magnitude if both these effects are at play in our systems. Here, we speculate briefly on the observational consequences of those modifications, with an eye toward clear tests that could validate or dispute their importance in explaining our observed trends.

Perhaps the most straightforward of these possibilities to test is the presence of buried star formation. From the perspective of our SED fitting, only J1157+0132 is better fit in the IR by the set of priors that favor buried star formation (and even then, only a $\sim0.14$ dex boost to the IR luminosity is needed to fit the 250 $\mu$m data). In all remaining galaxies, there is very little appreciable improvement in the fits, and only the choice of prior impacts the star formation rate. Outside of SED fitting, direct measurements of the star formation rate in \squiggle galaxies have come from rest-optical tracers, with the most rigorous constraints coming from weak $H\alpha$ emission \citep{Zhu2025}. However, $H\alpha$ is highly susceptible to dust attenuation. If the effective $A_V$ in $H_\mathrm{II}$ regions is $\gtrsim 3$, as allowed in our Buried SF Allowed fits, the true $H\alpha$ luminosity, and therefore star formation rates, could be higher than is inferred in that work by a factor of 10. This degeneracy can be broken with longer wavelength observations of instantaneous star formation rate tracers like Paschen and Brackett series emission, as the rest-NIR is far less susceptible to the effects of dust attenuation than the rest optical. Such observations of low-z post-starbursts found that there is little evidence from instantaneous SFR tracers of highly buried star formation \citep{Luo2022,French2023a}. However, no such observations have been presented for massive, CO-luminous post-starburst systems like those in \squigglecomma. 

Addressing the conversion from CO luminosity to $H_2$ mass remains important given that varying assumptions can lead to a factor of 5 difference in the inferred star forming material \citep{Bolatto2013}. One potential way forward would be instead estimating the molecular gas mass from the dust continuum, but different choices in dust emissivity \citep[e.g.,][]{Dunne2003, Draine2007} can easily induce factor-of-2 changes in the dust mass, as can shifts of $\sim20$ K in dust temperature. Furthermore, the gas-to-dust ratio, which would underpin this conversion, is itself found to have order-of-magnitude level uncertainty in quiescent systems in both simulations \citep{Whitaker2021b} and observations \citep{Spilker2025}, making any comparison from dust to gas just as systematically uncertain. That said, the application of some very standard choices of $T_{dust}=25$ K, \cite{Dunne2003} emissivity, and the $\delta_{GDR}=100$ (gas-to-dust ratio) to our four most luminous CO systems (all of which are detected in the continuum at 2mm) suggests that $\alpha_{CO}=2-3.5$, nowhere near the factor-of-5 reduction that would be obtained by invoking a ULIRG-like $\alpha_{CO}=0.8$. This same preference for Milky Way-like $\alpha_{CO}$ was found for the two \squiggle systems studied in \cite{Donofrio2025} under these same assumptions. However, all comparisons under standard $\delta_{GDR}$ assumptions are currently highly uncertain given that quiescent galaxies clearly depart from the typical relations between these quantities \citep{Spilker2025}.

Addressing the importance of mechanical gas removal is also fairly straightforward. \cite{Spilker2022} and \cite{Donofrio2025} have already shown that in many CO-luminous galaxies, as much as 50\% of the CO luminosity lies at distances $r>10$ kpc, and that gas, in the harsh environment of the circumgalactic medium, could potentially be disrupted without the need to form additional stars. Future high-resolution studies could better understand the bulk kinematics and spatial distribution of potential star forming gas. Absorption line studies could also search for evidence of outflowing material, which could have been launched by supernovae or AGN activity during the starburst and may play a role in the removal of the last of the ISM \citep[e.g.,][]{Geach2018, Belli2023, Sun2025_PSBoutflow}.

At present, there is little data-driven pressure toward any modification to the star formation rate (from buried star formation) or the $H_2$ mass (from a systematic lowering of $\alpha_{CO}$ or mechanical removal of star forming material) in the depletion time calculations for \squiggle post-starburst galaxies. However, with current data, it is also not possible to rule out either of these modifications. Given that without them, there is no way to interpret the trend in Figure \ref{fig:tpsb_LCO} as an evolutionary trend, future observations that can test these assumptions are crucial to understanding the future evolution of these perplexing CO-luminous systems.

\subsection{A final possibility: rejuvenation removes the need for rapid depletion} \label{subsec:rejuvenation}

Finally, we acknowledge that the assumption that underpins the aforementioned discussion--that the trend in Figure \ref{fig:tpsb_LCO} should be interpreted as an evolutionary sequence of CO luminosity with time since quenching--may be wrong. It remains a distinct possibility, regardless of exactly what the present day star formation rate of \squiggle galaxies is, that the young, gas rich galaxies in our sample are simply in a temporary lull where the combination of dust and a bursty star formation history post-merger make them temporarily appear quiescent (an effect that is seen in simulations, see Cenci et al. in preparation). A small number ($\sim25\%$) of the young ($t_\mathrm{PSB}<200$ Myr) \squiggle galaxies are not detected in CO(2-1), and even among CO-detected galaxies, there is an order of magnitude spread in implied molecular gas mass. The CO-poor (or undetected) \squiggle galaxies may represent ``true" post-starburst galaxies that are entering into quiescence, while the gas rich galaxies have not yet finished their life as star forming galaxies and will experience one or more additional periods of star formation before quenching in actuality.

There is some qualitative evidence that this might be the case if we take the ansatz that all massive \squiggle post-starburst galaxies have their starburst and quenching induced by a gas rich major merger, and that the absence of clear tidal features signals that the merger occurred further in the past. Of the most CO-luminous galaxies in our sample, nine present very bright tidal features that were flagged in \cite{Verrico2023}, and the remaining one (J0909+0108) was flagged as a close pair--though it too exhibits a clear tidal feature to the east. These galaxies could be those where a first burst has occurred, imparting enough feedback (stellar and AGN) into the remaining material to lower the star formation efficiency temporarily. These galaxies will eventually begin bursting again, and only after this will they enter the gas-poor post-starburst population, with considerably fainter tidal features owing to the additional time that has passed since the initial merger. In low-z systems, studies have found that AGN activity occurs almost immediately after coalescence, while an elevated post-starburst fraction occurs 160-480 Myr \textit{after} the merger \citep{Ellison2024_merger, Ellison2025_AGN}. A similar sequence could occurring in the \squiggle sample, and our CO luminous galaxies may be fated to become starbursts again on very short ($\sim$100 Myr) timescales.

Quantitatively addressing the impact of rejuvenation within the \squiggle sample is more difficult, because the completeness of our selection and the SDSS targeting as a function of age, burst mass, etc. is not well quantified. Self-consistently characterizing the timescales of quiescence would require a full census of star-forming and quiescent populations above a mass limit. With such a sample, one could imagine a detailed accounting of the number density of quiescent galaxies as a function of age to test whether galaxies indeed experience multiple periods where they appear quiescent before re-joining the star forming population. Such a signal would manifest in a pile up of sources at young ($t<200$ Myr) age, where sources like the gas rich galaxies in \squiggle could live once after an initial burst and again after they exhaust their final gas supply. Future surveys like the Prime Focus Spectrograph Galaxy Evolution Survey \citep{Greene2022} may be able to make significant strides here, with mass complete spectroscopic samples of galaxies at cosmic noon where rapid quenching must be happening in significant numbers. However, future detailed studies of the ISM in \squiggle galaxies may also offer clues about whether rejuvenation is likely in individual galaxies. 

\section{Conclusions} \label{sec:conclusions}

In this work, we perform a systematic study of CO(2-1) in 50 massive ($\log(M_\star/M_\odot)\sim11.2$) post-starburst galaxies at $z\sim0.7$, finding that a large fraction (27/50) are detected at the 3$\sigma$ level. We then set out to understand why these galaxies still host large molecular gas reservoirs despite appearing to be quiescent in the rest-optical \citep{Suess2022a, Zhu2025}. 

First, we address the question of whether these galaxies are still forming stars at an elevated rate behind optically thick dust. To do so, we perform new spectrophotometric fitting incorporating WISE W3/W4 photometry that was not included in previous fits, in addition to Herschel/SPIRE 250 $\mu$m photometry for a small number of sources that are detected. While we find that the full panchromatic SEDs of these galaxies can be well replicated without the need for buried star formation (see Figures \ref{fig:demo_IR_bright}, \ref{fig:demo_IR_faint}, and \ref{fig:model_comparisons}), allowing for optically thick birth cloud dust can increase inferred star formation rates by $\sim0.5$ dex. This shift can place some of these galaxies onto the star forming main sequence as measured in \cite[][see Figure \ref{fig:SFMS}]{Leja2022}, but there is very little room for any of them to be in the midst of active starbursts in our modeling framework.

We then address the time evolution of the CO luminosity. Previous findings indicate the CO luminosity drops by an order of magnitude in the few hundred Myr after quenching \citep[][see Figure \ref{fig:tpsb_LCO}]{French2015, Bezanson2022a}. We find that this implies depletion times of $\lesssim100$ Myr, and we set out to understand the conditions that could lower their CO luminosities by an order of magnitude on that timescale. We find that under the standard assumptions made in the literature about post-starburst galaxies \citep[that there is little to no star formation behind optically thick dust and that Milky Way assumptions about the CO thermalization and CO-to-$H_2$ conversions apply, see][]{French2015, Suess2017, Smercina2018, French2018a, Belli2021, Bezanson2022a, Zanella2023}, the depletion time is an order of magnitude longer than the necessary $\sim140$ Myr (see Figure \ref{fig:tdep}a).

We also find that the $\sim0.5$ dex boost from buried star formation is also not sufficient to match the observed $L_{CO(2-1)}$ versus age trend (see Figure \ref{fig:tdep}b), and the assumption of a lower starburst $\alpha_{CO}$ alone similarly can achieve the necessary depletion times (see Figure \ref{fig:tdep}c). Together, some amount of contribution from either of these mechanisms can lower the depletion times by more than an order of magnitude, but even still, the target depletion time of $\sim100$ Myr remains out of reach and cleanly explaining the age versus gas fraction trend as an evolution from gas-rich to gas-poor remains difficult.

At present, the evolutionary state of CO-luminous \squiggle galaxies remains unclear. Is there moderate star formation occurring behind optically thick dust, or are even the CO-luminous galaxies consistent with lying well below the star forming main sequence? Are they host to $>10^{10}$ $M_\odot$ molecular gas reservoirs, or is their molecular gas mass considerably overestimated due to the application of a Milky Way $\alpha_{CO}$ value?  If the former is the case, what is stabilizing these large molecular gas reservoirs against collapse? Is significant gas being removed in tidal driven stripping or outflows? And in any of the aforementioned scenarios, will these galaxies directly evolve directly into a quiescent population, or will they rejoin the star forming population before quenching permanently? 

Given the significant implications of all these possibilities to the evolutionary pathway that is thought to produce red-and-dead quiescent galaxies, further study of the ISM in these systems is needed. The formation of massive quiescent galaxies via rapid quenching process is empirically necessary, especially at high-z where it is essentially impossible to model observed spectra without extreme star formation that shuts off on short timescales \citep[e.g.,][]{Carnall2023b, Glazebrook2024, deGraaff2025_QG, Beverage2025}. Furthermore, the presence of sub-millimeter galaxies with strong Balmer breaks has demonstrated that the physical process that is acting in \squiggle galaxies is also likely at work at cosmic noon \citep[e.g.,][]{Cooper2025}. It is crucial that we use these lower-z analogues where detailed studies of gas and stars across the panchromatic SED are possible to understand the physics that drive the rapid cessation of star formation after starbursts.

\facilities{ALMA, SDSS, Subaru, WISE, Herschel}

\software{Astropy \citep{astropy2013, astropy2018, astropy2022},
Matplotlib \citep{Hunter:2007}, Flexible Stellar Population Synthesis \citep{Conroy2009, Conroy2010}, SEDPy \citep{sedpy2019}, Prospector \citep{Johnson2017, Leja2017, Johnson2021}}

\acknowledgements

DJS acknowledges Joel Leja for many helpful conversations regarding the implementation of dust emission in Prospector, Jared Siegel for helping to hack out modifications to star formation histories, and Dalya Baron for helpful discussions about FIR SED parameterization and Prospector prior choices. Support for this work was provided by The Brinson Foundation through a Brinson Prize Fellowship grant. DJS gratefully acknowldges support from NRAO Student Observing Support from NRAO-SOSPA8-008. KAS, JSS, and VRD gratefully acknowledge support from NSF-AAG\#2407954 and 2407955, and NRAO-SOSPA11-006.

This paper makes use of the following ALMA data: ADS/JAO.ALMA \#2016.1.01126.S, ADS/JAO.ALMA \#2017.1.01109.S, ADS/JAO.ALMA \#2021.1.01535.S, ADS/JAO.ALMA \#2021.1.00988.S,ADS/JAO.ALMA \#2021.1.00761.S, and ADS/JAO.ALMA \#2022.1.00604.S. ALMA is a partnership of ESO (representing its member states), NSF (USA) and NINS (Japan), together with NRC (Canada), NSTC and ASIAA (Taiwan), and KASI (Republic of Korea), in cooperation with the Republic of Chile. The Joint ALMA Observatory is operated by ESO, AUI/NRAO and NAOJ. The National Radio Astronomy Observatory is a facility of the National Science Foundation operated under cooperative agreement by Associated Universities, Inc. 

This publication makes use of data products from the Wide-field Infrared Survey Explorer, which is a joint project of the University of California, Los Angeles, and the Jet Propulsion Laboratory/California Institute of Technology, funded by the National Aeronautics and Space Administration.

Funding for the Sloan Digital Sky Survey IV has been provided by the Alfred P. Sloan Foundation, the U.S. Department of Energy Office of Science, and the Participating Institutions. SDSS acknowledges support and resources from the Center for High-Performance Computing at the University of Utah. The SDSS web site is www.sdss4.org.

SDSS is managed by the Astrophysical Research Consortium for the Participating Institutions of the SDSS Collaboration including the Brazilian Participation Group, the Carnegie Institution for Science, Carnegie Mellon University, Center for Astrophysics | Harvard \& Smithsonian (CfA), the Chilean Participation Group, the French Participation Group, Instituto de Astrofísica de Canarias, The Johns Hopkins University, Kavli Institute for the Physics and Mathematics of the Universe (IPMU) / University of Tokyo, the Korean Participation Group, Lawrence Berkeley National Laboratory, Leibniz Institut für Astrophysik Potsdam (AIP), Max-Planck-Institut für Astronomie (MPIA Heidelberg), Max-Planck-Institut für Astrophysik (MPA Garching), Max-Planck-Institut für Extraterrestrische Physik (MPE), National Astronomical Observatories of China, New Mexico State University, New York University, University of Notre Dame, Observatório Nacional / MCTI, The Ohio State University, Pennsylvania State University, Shanghai Astronomical Observatory, United Kingdom Participation Group, Universidad Nacional Autónoma de México, University of Arizona, University of Colorado Boulder, University of Oxford, University of Portsmouth, University of Utah, University of Virginia, University of Washington, University of Wisconsin, Vanderbilt University, and Yale University.

The Herschel spacecraft was designed, built, tested, and launched under a contract to ESA managed by the Herschel/Planck Project team by an industrial consortium under the overall responsibility of the prime contractor Thales Alenia Space (Cannes), and including Astrium (Friedrichshafen) responsible for the payload module and for system testing at spacecraft level, Thales Alenia Space (Turin) responsible for the service module, and Astrium (Toulouse) responsible for the telescope, with in excess of a hundred subcontractors.

The Hyper Suprime-Cam (HSC) collaboration includes the astronomical communities of Japan and Taiwan, and Princeton University. The HSC instrumentation and software were developed by the National Astronomical Observatory of Japan (NAOJ), the Kavli Institute for the Physics and Mathematics of the Universe (Kavli IPMU), the University of Tokyo, the High Energy Accelerator Research Organization (KEK), the Academia Sinica Institute for Astronomy and Astrophysics in Taiwan (ASIAA), and Princeton University. Funding was contributed by the FIRST program from Japanese Cabinet Office, the Ministry of Education, Culture, Sports, Science and Technology (MEXT), the Japan Society for the Promotion of Science (JSPS), Japan Science and Technology Agency (JST), the Toray Science Foundation, NAOJ, Kavli IPMU, KEK, ASIAA, and Princeton University. 

This paper makes use of software developed for the Large Synoptic Survey Telescope. We thank the LSST Project for making their code available as free software at  http://dm.lsst.org

The Pan-STARRS1 Surveys (PS1) have been made possible through contributions of the Institute for Astronomy, the University of Hawaii, the Pan-STARRS Project Office, the Max-Planck Society and its participating institutes, the Max Planck Institute for Astronomy, Heidelberg and the Max Planck Institute for Extraterrestrial Physics, Garching, The Johns Hopkins University, Durham University, the University of Edinburgh, Queen’s University Belfast, the Harvard-Smithsonian Center for Astrophysics, the Las Cumbres Observatory Global Telescope Network Incorporated, the National Central University of Taiwan, the Space Telescope Science Institute, the National Aeronautics and Space Administration under Grant No. NNX08AR22G issued through the Planetary Science Division of the NASA Science Mission Directorate, the National Science Foundation under Grant No. AST-1238877, the University of Maryland, and Eotvos Lorand University (ELTE) and the Los Alamos National Laboratory.

Based in part on data collected at the Subaru Telescope and retrieved from the HSC data archive system, which is operated by Subaru Telescope and Astronomy Data Center at National Astronomical Observatory of Japan.

\appendix

\section{Full tables of observing parameters}

Here, we present Table \ref{tbl:alma_properties} and Table \ref{tbl:alma_measurements}, which contain the properties of our ALMA observations and the measured CO(2-1) and continuum properties of our sources. 

\section{Comparison of ``time since quenching" and ``post-starburst time", and alternate definitions} \label{sec:timescale_comp}

Here, we compare our measured post-starburst times, which once again is defined as the lookback time where the galaxy formed 99\% of the total stellar mass it formed in the Gyr before observation, to the time since quenching measured in \cite{Suess2022a}. The \cite{Suess2022a} time since quenching relied on taking a derivative of the star formation history, and is therefore sensitive to the parameterization and also whether the galaxy dropped enough to meet the specific definition. This makes the measurement impossible for galaxies that do not fully quench, which becomes increasingly more likely the star formation histories inferred under our Buried SF Allowed priors.

\begin{figure*}
    \centering
    \includegraphics[width=\textwidth]{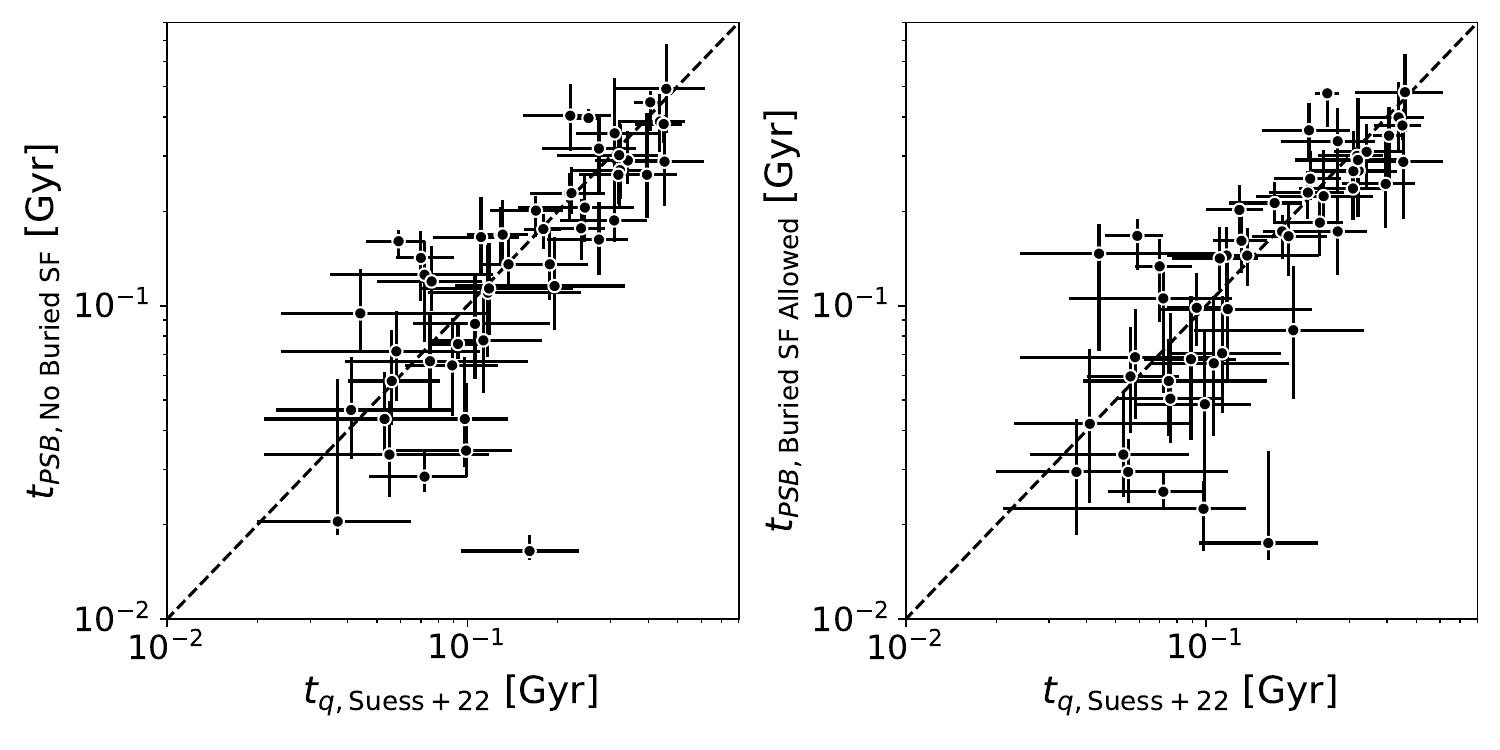}
    \caption{A comparison between the time since quenching ($t_q$) measured in \cite{Suess2022a} and the post-starburst time ($t_\mathrm{PSB}$) we measure in this work, for the No Buried SF prior (left) and Buried SF Allowed prior (right). For both sets of the fits, these quantities scatter around the 1:1 line with minimal scatter.}
    \label{fig:timescale_comp}
\end{figure*}

In Figure \ref{fig:timescale_comp}, we compare these metrics for the star formation histories derived in our No Buried SF and Buried SF Allowed priors, finding very good agreement with the \cite{Suess2022a} time since quenching in both cases. The largest outlier, which is $\sim150$ Myr old in the \cite{Suess2022a} fits but is closer to 15 Myr old by our new metric, is J0910+0218 (see Figure \ref{fig:demo_IR_medium}). This difference in age reflects an actual change in the star formation history of the source relative to the \cite{Suess2022a} fits; the inclusion of mid- and far-IR photometry necessitated a younger age to properly capture the full SED in this source. Thus, we propose that this $t_\mathrm{PSB}$ metric, which has the added advantage of being very easy to compare to simulations, can serve as a good proxy for the time since quenching.

That said, our definition of $t_{PSB}$ is not without problems. As burst gets older, the measure becomes increasingly susceptible to being pulled to younger ages by low lying levels of star formation. As such, it may also be desirable to compute other moments of the star formation history that are more robust to this effect by tracing star formation on longer timescales. Thus, we compute a measure of the post-starburst time, $t_{\mathrm{PSB},90}$, that is identical in definition to our fiducial $t_\mathrm{PSB}$, but instead measuring the lookback time where 90\% of the stars formed in the last Gyr. This metric tends to measure a value that is closer to the peak of the burst, especially in younger galaxies. 

Finally, we seek to measure a quantity which is as close as possible to the quenching time defined in \cite{Suess2022a} but which is still possible to calculate for galaxies which do not drop significantly enough to be classified by a derivative. For this, we adopt a metric $t_{q,10^{-10} \  \mathrm{yr}^{-1}}$, which we define as the lookback time where the star formation rate first exceeds $M_\star \times 10^{-10} \ \mathrm{yr}^{-1}$. If the galaxy is already measured to be instantaneously forming stars above this rate, the metric is measured as 1 Myr. This metric has the disadvantage of being scale variant, insofar as we assume an arbitrary cutoff in the specific star formation rate that should in all likelihood evolve with redshift. However, is still serves as a useful measure that can trace the last time the galaxy was forming stars at anywhere near main sequence star formation rates.

\begin{figure*}
    \centering
    \includegraphics[width=0.49\textwidth]{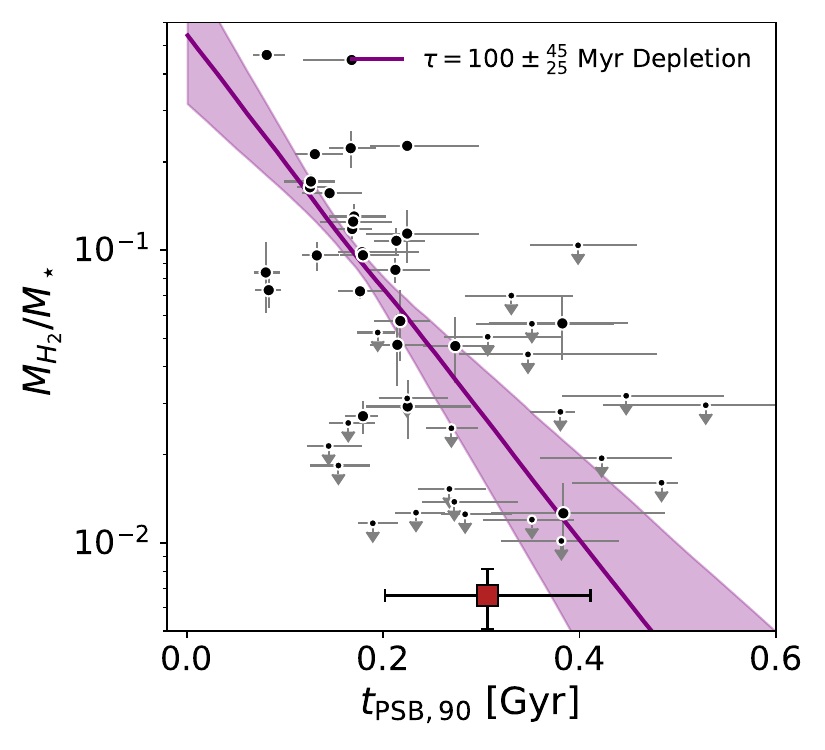}
    \includegraphics[width=0.49\textwidth]{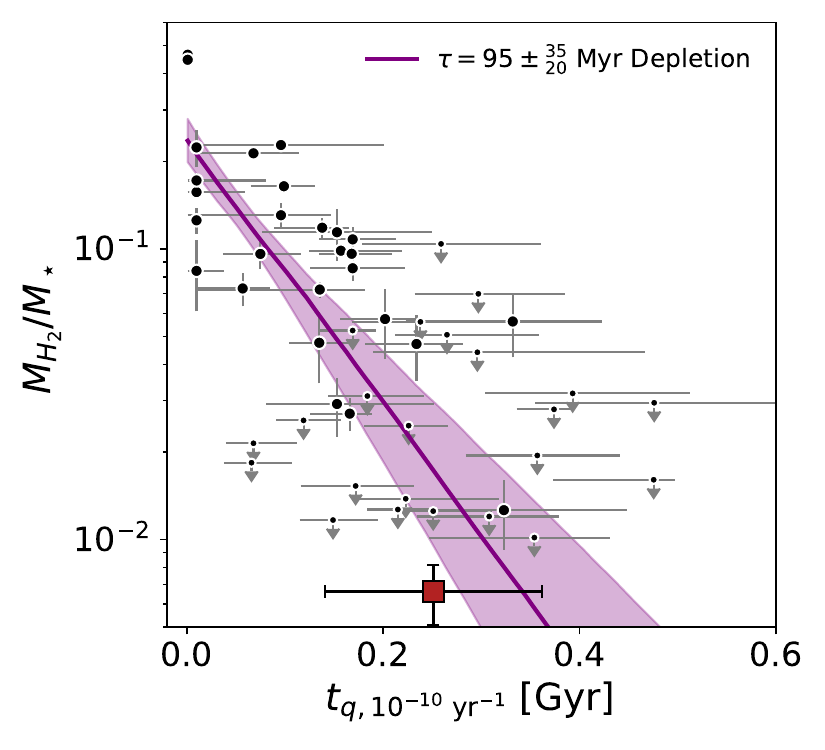}
    \caption{The gas fraction versus age, using our alternate metrics of age defined in Appendix \ref{sec:timescale_comp}. While the specific age we measure differs depending on the timescale adopted, the qualitative behavior of the youngest galaxies being CO detected remains, and the decay timescale we measure is robust within a factor of 2 with our fiducial $t_\mathrm{PSB}$ metric. }
    \label{fig:t_test_Mgas}
\end{figure*}

In Figure \ref{fig:t_test_Mgas}, we present the same gas fraction vs age plot as in Figure \ref{fig:tpsb_LCO}, but instead using these different age metrics measured from our Buried SF Allowed fits. We use a similar fitting procedure to measure the decay timescale, and find we measure a timescale for both these parameters that are 1$\sigma$ consistent with the measurements using our fiducial $t_\mathrm{PSB}$. As such, the finding of decay that is too rapid to be produced by even the combination of buried star formation and a significantly lower $H_2$ mass is robust to the definition of age.

% While the specific timescale that we infer for the decay in the gas fraction changes by a factor of $\sim2$ relative to the timescale we measure with our fiducial $t_{\mathrm{PSB}}$ (especially because the uncertainty in the age measure, which is not being accounted for in our simple fitting, is much larger in these metrics), the qualitative behavior where the youngest galaxies are far more likely to be CO(2-1) detected than the older ones remains. In Table \ref{tbl:different_age}, we present these different age metrics for our sample, for both the No Buried SF and Buried SF Allowed fits. 

\begin{deluxetable*}{cccccccc}
\tablecaption{Properties of the ALMA observations. \label{tbl:alma_properties}}
\tablehead{\colhead{ID} & \colhead{RA} & \colhead{Dec} & \colhead{z} & \colhead{Program ID\tablenotemark{a}} & \colhead{Observing Date} & \colhead{Integration Time} & \colhead{Angular Resolution} \\
\colhead{} & \colhead{[deg]} & \colhead{[deg]} & \colhead{} & \colhead{} & \colhead{} & \colhead{[s]} & \colhead{[$''$]} }
\startdata
J0131+0034 & 22.885654 & $+$0.577446 & 0.618268 & 2021.1.01535.S & 3/26/22 & 2999.8 & 2.1 \\
J0153-0207 & 28.265102 & $-$2.126556 & 0.66585 & 2021.1.01535.S/2022.1.00604.S & 12/22/2022, 4/30/2022 & 7801.9 & 1.2 \\
J0221-0646 & 35.363172 & $-$6.778846 & 0.661296 & 2021.1.01535.S & 3/22/22 & 2993.8 & 2.2 \\
J0224+0015 & 36.178065 & $+$0.253695 & 0.684893 & 2021.1.01535.S & 3/26/22 & 2963.5 & 2.1 \\
J0224-0034 & 36.126895 & $-$0.571105 & 0.743342 & 2022.1.00604.S & 12/24/22 & 2993.8 & 1.5 \\
J0224-0630 & 36.039585 & $-$6.511067 & 0.751483 & 2021.1.01535.S & 1/28/22 & 2993.8 & 1.4 \\
J0227-0225 & 36.99331 & $-$2.430015 & 0.61386 & 2021.1.00988.S & 1/28/22 & 5927.0 & 1.4 \\
J0232-0331 & 38.209972 & $-$3.52757 & 0.738828 & 2022.1.00604.S & 1/26/22 & 5927.0 & 1.4 \\
J0237+0123 & 39.265248 & $+$1.390005 & 0.709438 & 2021.1.01535.S/2022.1.00604.S & 9/11/2022, 12/27/2022 & 5999.6 & 1.2 \\
J0240-0148 & 40.054477 & $-$1.81099 & 0.689513 & 2021.1.01535.S & 1/27/22 & 2993.8 & 1.4 \\
J0851+0244 & 132.76778 & $+$2.742179 & 0.659483 & 2022.1.00604.S & 12/17/22 & 2999.8 & 1.4 \\
J0907+0423 & 136.99116 & $+$4.384259 & 0.663469 & 2021.1.00988.S & 4/21/22 & 5806.1 & 1.5 \\
J0909-0108 & 137.47026 & $-$1.134891 & 0.702088 & 2021.1.01535.S & 1/27/22 & 2993.8 & 1.4 \\
J0910+0218 & 137.61906 & $+$2.309311 & 0.769447 & 2022.1.00604.S & 12/31/22 & 2993.8 & 1.6 \\
J0940-0008 & 145.1991 & $-$0.138624 & 0.618503 & 2022.1.00604.S & 12/28/22 & 2999.8 & 1.4 \\
J1017-0003 & 154.30202 & $-$0.061552 & 0.596369 & 2022.1.00604.S & 12/27/22 & 2993.8 & 1.3 \\
J1040+0223 & 160.04608 & $+$2.391646 & 0.680527 & 2022.1.00604.S & 1/3/23 & 2993.8 & 1.4 \\
J1042+0500 & 160.71635 & $+$5.010805 & 0.626593 & 2022.1.00604.S & 12/31/22 & 2993.8 & 1.4 \\
J1046+0123 & 161.7372 & $+$1.399393 & 0.764902 & 2022.1.00604.S & 1/19/23 & 2993.8 & 1.0 \\
J1114+0115 & 168.60707 & $+$1.265318 & 0.748912 & 2022.1.00604.S & 1/19/23 & 2993.8 & 1.0 \\
J1141-0109 & 175.38035 & $-$1.151463 & 0.657525 & 2022.1.00604.S & 1/2/23 & 2993.8 & 1.5 \\
J1142+0006 & 175.53723 & $+$0.105861 & 0.593473 & 2021.1.00988.S & 5/8/22 & 5745.6 & 1.3 \\
J1148+0204 & 177.1627 & $+$2.069523 & 0.596461 & 2022.1.00604.S & 12/31/22 & 2993.8 & 1.4 \\
J1157+0132 & 179.48879 & $+$1.537532 & 0.755935 & 2021.1.00988.S & 5/9/22 & 2993.8 & 1.5 \\
J1211+0240 & 182.90725 & $+$2.673695 & 0.589034 & 2022.1.00604.S & 10/16/22 & 2993.8 & 1.7 \\
J1222+0342 & 185.65243 & $+$3.711715 & 0.679837 & 2021.1.01535.S & 6/3/22 & 2993.8 & 1.0 \\
J1240-0057 & 190.18837 & $-$0.951925 & 0.795872 & 2022.1.00604.S & 1/19/23 & 2993.8 & 1.1 \\
J1244+0248 & 191.00193 & $+$2.807904 & 0.625788 & 2021.1.01535.S & 8/30/22 & 2993.8 & 0.9 \\
J1332+0256 & 203.11942 & $+$2.933643 & 0.698313 & 2021.1.01535.S/2022.1.00604.S & 8/28/2022, 12/27/2022 & 5987.5 & 1.2 \\
J1416+0255 & 214.00084 & $+$2.924476 & 0.760227 & 2022.1.00604.S & 12/27/22 & 2993.8 & 1.6 \\
J1436+0447 & 219.16637 & $+$4.795968 & 0.633855 & 2021.1.00761.S & 8/28/22 & 5987.5 & 0.9 \\
J1437+0311 & 219.43569 & $+$3.19169 & 0.667036 & 2021.1.01535.S/2022.1.00604.S & 8/28/22 & 5987.5 & 1.2 \\
J1444-0006 & 221.03227 & $-$0.106326 & 0.700658 & 2021.1.01535.S & 4/6/22 & 2999.8 & 2.3 \\
J1449+0206 & 222.33699 & $+$2.101912 & 0.743967 & 2022.1.00604.S & 12/26/22 & 2999.8 & 1.7 \\
J1455-0048 & 223.93174 & $-$0.8092 & 0.625287 & 2021.1.01535.S & 4/6/22 & 2999.8 & 2.1 \\
J2213-0050 & 333.45469 & $-$0.833433 & 0.66203 & 2021.1.01535.S/2022.1.00604.S & 4/06/2022, 12/20/2022 & 4814.2 & 1.5 \\
J2232+0007 & 338.16968 & $+$0.131622 & 0.722158 & 2021.1.01535.S/2022.1.00604.S & 9/10/2022, 12/29/2022 & 5987.5 & 1.2 \\
J2310-0047 & 347.59742 & $-$0.798654 & 0.737807 & 2021.1.00988.S & 1/25/22 & 5999.6 & 1.4 \\
\enddata
\tablenotetext{a}{Some sources were observed in multiple cycles due to not being flagged as duplicates when they were observed between submission of proposals and the beginning of the new cycle. For these sources, we list both program IDs and combine all data our imaging.}
\end{deluxetable*}

\begin{deluxetable*}{ccccccc}
\tablecaption{CO(2-1) and continuum measurements in a 2$''$ aperture. All non-detections are listed as 3$\sigma$ upper limits. For sources in the Herschel/SPIRE $250$ $\mu$m source catalog, we also list their flux density. \label{tbl:alma_measurements}}
\tablehead{\colhead{ID} & \colhead{SdvCO(2–1)} & \colhead{LCO(2-1)} & \colhead{$M_{H_2}$\tablenotemark{a}} & \colhead{$f_{2\mathrm{mm}}$} & \colhead{$f_{250\mu\mathrm{m}}$} \\
\colhead{} & \colhead{[J km $\mathrm{s^{-1}}$]} & \colhead{[$10^{9}$ J km $\mathrm{s^{-1} \ pc^{2}}$]} & \colhead{[$10^{10} \ M_\odot$]} & \colhead{[$\mu$Jy]} & \colhead{[mJy]} }
\startdata
J0131+0034 & $0.05 \pm 0.01$ & $0.26 \pm 0.07$ & $0.11 \pm 0.03$ & $<26.55$ & -- \\
J0153-0207 & $0.38 \pm 0.04$ & $2.37 \pm 0.23$ & $0.95 \pm 0.09$ & $<30.91$ & -- \\
J0221-0646 & $<0.09$ & $<0.53$ & $<0.21$ & $<25.81$ & -- \\
J0224+0015 & $<0.12$ & $<0.76$ & $<0.31$ & $<27.29$ & -- \\
J0224-0034 & $0.51 \pm 0.06$ & $3.99 \pm 0.45$ & $1.59 \pm 0.18$ & $146.14 \pm 15.17$ & -- \\
J0224-0630 & $0.63 \pm 0.06$ & $5.0 \pm 0.51$ & $2.0 \pm 0.2$ & $<43.46$ & -- \\
J0227-0225 & $1.09 \pm 0.04$ & $5.79 \pm 0.22$ & $2.32 \pm 0.09$ & $<36.84$ & -- \\
J0232-0331 & $<0.14$ & $<1.09$ & $<0.43$ & $<40.54$ & -- \\
J0237+0123 & $0.62 \pm 0.06$ & $4.38 \pm 0.44$ & $1.75 \pm 0.17$ & $<43.94$ & -- \\
J0240-0148 & $<0.25$ & $<1.68$ & $<0.67$ & $<72.5$ & -- \\
J0851+0244 & $0.61 \pm 0.06$ & $3.77 \pm 0.39$ & $1.51 \pm 0.16$ & $65.49 \pm 18.65$ & -- \\
J0907+0423 & $2.32 \pm 0.03$ & $14.41 \pm 0.17$ & $5.76 \pm 0.07$ & $93.69 \pm 10.64$ & -- \\
J0909-0108 & $1.0 \pm 0.14$ & $6.98 \pm 1.01$ & $2.79 \pm 0.4$ & $<84.86$ & -- \\
J0910+0218 & $1.82 \pm 0.04$ & $15.28 \pm 0.37$ & $6.11 \pm 0.15$ & $42.91 \pm 11.76$ & $50.3 \pm 13.8$ \\
J0940-0008 & $<0.2$ & $<1.1$ & $<0.44$ & $<48.36$ & -- \\
J1017-0003 & $<0.13$ & $<0.65$ & $<0.26$ & $<43.11$ & -- \\
J1040+0223 & $<0.16$ & $<1.05$ & $<0.42$ & $<39.51$ & -- \\
J1042+0500 & $0.2 \pm 0.05$ & $1.12 \pm 0.29$ & $0.45 \pm 0.11$ & $<34.69$ & -- \\
J1046+0123 & $<0.35$ & $<2.87$ & $<1.15$ & $<76.3$ & -- \\
J1114+0115 & $0.5 \pm 0.14$ & $3.99 \pm 1.09$ & $1.6 \pm 0.44$ & $<86.31$ & -- \\
J1141-0109 & $<0.14$ & $<0.87$ & $<0.35$ & $<39.21$ & -- \\
J1142+0006 & $2.27 \pm 0.05$ & $11.22 \pm 0.24$ & $4.49 \pm 0.1$ & $77.11 \pm 12.65$ & $43.2 \pm 13.1$ \\
J1148+0204 & $<0.13$ & $<0.66$ & $<0.26$ & $<39.16$ & -- \\
J1157+0132 & $3.49 \pm 0.05$ & $28.26 \pm 0.4$ & $11.3 \pm 0.16$ & $102.37 \pm 13.71$ & $84.3 \pm 19.6$ \\
J1211+0240 & $0.24 \pm 0.06$ & $1.18 \pm 0.27$ & $0.47 \pm 0.11$ & $<36.85$ & -- \\
J1222+0342 & $0.3 \pm 0.08$ & $1.96 \pm 0.54$ & $0.78 \pm 0.21$ & $<58.16$ & -- \\
J1240-0057 & $<0.37$ & $<3.34$ & $<1.34$ & $<72.29$ & -- \\
J1244+0248 & $0.86 \pm 0.18$ & $4.73 \pm 0.97$ & $1.89 \pm 0.39$ & $<55.39$ & -- \\
J1332+0256 & $0.23 \pm 0.06$ & $1.61 \pm 0.4$ & $0.64 \pm 0.16$ & $47.93 \pm 12.06$ & -- \\
J1416+0255 & $<0.13$ & $<1.07$ & $<0.43$ & $<36.56$ & -- \\
J1436+0447 & $1.23 \pm 0.07$ & $6.94 \pm 0.42$ & $2.78 \pm 0.17$ & $<61.1$ & -- \\
J1437+0311 & $<0.14$ & $<0.88$ & $<0.35$ & $<38.94$ & -- \\
J1444-0006 & $0.3 \pm 0.04$ & $2.06 \pm 0.27$ & $0.82 \pm 0.11$ & $<39.15$ & -- \\
J1449+0206 & $0.24 \pm 0.06$ & $1.87 \pm 0.51$ & $0.75 \pm 0.2$ & $<44.86$ & -- \\
J1455-0048 & $<0.19$ & $<1.06$ & $<0.42$ & $<38.3$ & -- \\
J2213-0050 & $<0.16$ & $<0.97$ & $<0.39$ & $<34.7$ & -- \\
J2232+0007 & $<0.21$ & $<1.52$ & $<0.61$ & $<35.79$ & -- \\
J2310-0047 & $1.13 \pm 0.04$ & $8.7 \pm 0.34$ & $3.48 \pm 0.14$ & $<39.28$ & -- \\
\enddata
\tablenotetext{a}{Assuming $r_{21}=1.0$ and $\alpha_{CO}=4.0$}
\end{deluxetable*}

\begin{deluxetable*}{ccc}
\tabletypesize{}
\tablecaption{Alternate measures of the age of the starburst for these galaxies. The full table is available in a machine readable form. \label{tbl:different_age}}
\tablehead{\colhead{ID} & \colhead{$t_{\mathrm{PSB},90}$} & \colhead{$t_{q,10^{-10} \  \mathrm{yr}^{-1}}$} \\
\colhead{} & \colhead{[Gyr]} & \colhead{[Gyr]}
}
\startdata
\textbf{No Buried Star Formation:} & & \\
J1157+0132 & 0.117$\pm^{0.03}_{0.018}$ & 0.01$\pm^{0.041}_{0.009}$ \\
J0910+0218 & 0.084$\pm^{0.017}_{0.017}$ & 0.001$\pm^{0.009}_{0.0}$ \\
J1141-0109 & 0.272$\pm^{0.028}_{0.023}$ & 0.225$\pm^{0.052}_{0.034}$ \\
... & ... \\
\textbf{Buried Star Formation Allowed:} & & \\
J1157+0132 & 0.168$\pm^{0.053}_{0.049}$ & 0.001$\pm^{0.0}_{0.0}$ \\
J0910+0218 & 0.082$\pm^{0.019}_{0.014}$ & 0.001$\pm^{0.0}_{0.0}$ \\
J1141-0109 & 0.284$\pm^{0.048}_{0.024}$ & 0.251$\pm^{0.06}_{0.038}$ \\
... & ... \\
\enddata

\end{deluxetable*}

\bibliography{SQuIGGLE-Gas}

\end{document}